\documentclass[prd,aps,showpacs,nofootinbib,showkeywords,eqsecnum,twocolumn]{revtex4-1}

\usepackage{graphicx,color,amsmath,amsxtra}
\usepackage{epsf}
\usepackage{amssymb}
\usepackage{enumerate}
\usepackage{hhline}
\usepackage{array}
\usepackage{tabularx}
\usepackage{hangcaption}
\usepackage[unicode]{hyperref}
\usepackage{graphicx}              
\usepackage{epstopdf}

\hypersetup{
  colorlinks   = true, 
  urlcolor     = blue, 
  linkcolor    = blue, 
  citecolor   = blue 
}
  
\begin{document}
\pagestyle{plain}
\pagestyle{myheadings}

\title{Higher dimensional nonlinear massive gravity}
\author{Tuan Q. Do}
\email{tuanqdo@vnu.edu.vn} \email{tuanqdo.py97g@nctu.edu.tw}
\affiliation{Faculty of Physics, VNU University of Science, Vietnam National University,  Hanoi, Vietnam}
\date{\today }

\begin{abstract}
Inspired by a recent ghost-free nonlinear massive gravity in four-dimensional spacetime, we study its higher dimensional scenarios.  As a result, we are able to show the constant-like behavior of massive graviton terms for some well-known metrics such as the Friedmann-Lemaitre-Robertson-Walker, Bianchi type I, and  Schwarzschild-Tangherlini-(A)dS  metrics in a specific five-dimensional nonlinear massive gravity under an assumption  that its fiducial metrics are compatible with physical ones. In addition, some simple cosmological solutions of the five-dimensional massive gravity will be figured out consistently. 

\end{abstract}

%


\maketitle
\section{Introduction} \label{sec1}
Recently, de Rham, Gabadadze, and Tolley (dRGT) have successfully constructed a nonlinear massive gravity ~\cite{RGT} as a generalization of massive gravity proposed by Fierz and Pauli   long time ago ~\cite{FP,NAH}. More importantly, the dRGT theory with an arbitrary fiducial (or reference) metric has been proved  to be free of the so-called Boulware-Deser (BD) ghost ~\cite{BD}, which is associated with the sixth mode in graviton coming from nonlinear levels ~\cite{Arnowitt:1962hi}, 
by different approaches ~\cite{proof}. It is known that in the ADM language ~\cite{Arnowitt:1962hi} the existence of the BD ghost is due to the absence of a Hamiltonian constraint. The success
of non-linear massive gravity is based on the fact that it admits such a constraint and an
associated secondary constraint eliminating the sixth degree of freedom corresponding to the BD ghost mode ~\cite{proof}. Thanks to the nonlinear and ghost-free properties, the dRGT theory has been expected to admit alternative solutions to some remaining problems in modern cosmology, e.g., the old cosmological constant problem and the dark energy problem. In addition, some exotic or novel results due to the existence of nonlinear graviton terms might emerge  in the context of the dRGT theory. In fact, many cosmological and physical aspects  of the dRGT theory have been investigated extensively. For recent interesting reviews of the development of the massive gravity theory, see Ref.  ~\cite{review}.

 In particular, ones showed in Ref.~\cite{FRW} that the nonlinear massive gravity with the Minkowski fiducial metric does not admit the spatially flat and closed Friedmann-Lemaitre-Robertson-Walker (FLRW) metrics, which are homogeneous and isotropic, as its cosmological solutions. However,  the spatially open  FLRW metric has been shown to exist in the dRGT theory  ~\cite{FGM}. In addition, some homogeneous but anisotropic (Bianchi type I) cosmological solutions  have been discovered recently ~\cite{bianchi-I,WFK}. Some well-known black holes such as the Schwarzschild, Kerr, and charged black holes  have also been claimed to appear in the context of nonlinear massive gravity ~\cite{KNT,RGT1,RGT2,TMN,kerr,Tolley:2015ywa,stability-BH,non-bidiagonal-BH,Cai:2014znn}. Surprisingly, the massive gravity has been applied into the  holographic descriptions of condensed matter systems ~\cite{Cai:2014znn,Vegh:2013sk}.

Inspired by the success of the nonlinear massive gravity theory,  a bi-gravity (bi-metric) theory that is also free of ghost has been proposed by Hassan and Rosen  ~\cite{SFH,soda,aoki}. In addition, ghost-free multi-metric theories have also been formulated in Ref.~\cite{NK}.  The main different point between the massive gravity and bi (multi)-gravity theories is that the fiducial metric  is non-dynamical in the massive theory but dynamical in bi (multi)-gravity theories. On the other hand, there have been some extensions of the dRGT theory, e.g., the quasi-dilaton model ~\cite{quasi} and its extension ~\cite{quasi-gen}, the mass-varying massive gravity ~\cite{varying}, the extended massive gravity ~\cite{extended}, the $f(R)$ nonlinear massive gravity ~\cite{fr}, the massive gravity with non-minimal coupling of matter ~\cite{non-minimal}, and the massive gravity nonlinear sigma model ~\cite{sigma}. To the best of our knowledge, there has been a paper ~\cite{higher-massive} discussing on a higher dimensional extension of the mass-varying massive gravity, where scalar fields are non-minimally coupled to a massive graviton. However,  specific higher dimensional graviton terms and their cosmological implications are not studied in this paper. It is worth noting that  a class of charged black hole solutions  of a higher dimensional massive gravity with a negative cosmological constant have been found in papers of Ref.~\cite{Cai:2014znn}. However,  these papers have focused only on the usual graviton terms, ${\cal L}_i$ ($i=2-4$), which were proposed in the four-dimensional dRGT theory ~\cite{RGT}.  Along this line of research, Ref. ~\cite{higher-bigravity} has analyzed the spectrum of the ghost-free bigravity theory in arbitrary (higher) dimensions. Note again that all of these papers have not discussed in detail scenarios involving additional higher dimensional graviton terms, e.g., ${\cal L}_5$ and ${\cal L}_6$ \cite{Bonifacio:2014rba}, which would not vanish in five-dimensional and six-dimensional spacetimes, respectively. 
Hence, studying the dRGT theory and its extensions with specific higher dimensional graviton terms in higher dimensions is physically important. For example, introducing the higher dimensional graviton terms into the corresponding higher dimensional scenarios of dRGT theory would modify results investigated in the previous papers, e.g., Refs.  ~\cite{Cai:2014znn,higher-massive,higher-bigravity}.

 In the present paper, therefore, we would like to construct explicit higher dimensional graviton terms of the (pure) nonlinear massive gravity and study their cosmological implications. As a result, our construction is based on an observation that the ghost-free graviton terms of the four-dimensional dRGT theory are indeed a consequence of the Cayley-Hamilton theorem ~\cite{CH-theorem}, which is associated with the definition of the determinant of square matrix.  In particular, by using the characteristic equation of square matrix, which is a consequence of the Cayley-Hamilton theorem, we are able to re-build up the four-dimensional ghost-free graviton terms, which have been shown to be free of the BD ghost.  Similarly, higher dimensional ghost-free graviton terms can also be constructed by applying the same technique used for the four-dimensional   graviton terms. In particular, we will explicitly define five-, six-, and  seven-dimensional graviton terms, ${\cal L}_5$,  ${\cal L}_6$, and ${\cal L}_7$, as clear demonstrations for our method. For heuristic reasons, we will discuss a five-dimensional nonlinear massive gravity involving an additional graviton term, ${\cal L}_5$.  In particular, we will examine whether some well-known metrics, e.g., Friedmann-Lemaitre-Robertson-Walker~\cite{5d-FRW}, Bianchi type I ~\cite{5d-bianchi}, and Schwarzschild-Tangherlini black hole ~\cite{5d-sch,5d-sch-stability}, exist in the five-dimensional nonlinear massive gravity. We will also check the constant-like behavior of massive graviton terms, which has been shown in the four-dimensional dRGT theory. 

This paper will be organized as follows: (i) A brief introduction and motivation of this research has been given in Sec. \ref{sec1}. (ii) Some basic details of four-dimensional nonlinear massive gravity theory will be presented in Sec. \ref{sec2} and  (iii) some specific higher dimensional graviton terms will be addressed in Sec. \ref{sec3}. (iv) The FLRW metric and its generalization, the Bianchi type I metric, will be studied in the five-dimensional nonlinear massive gravity in Sec. \ref{sec4} and Sec. \ref{sec5}, respectively. (v) Next, we will examine whether  the five-dimensional nonlinear massive gravity admits the Schwarzschild-Tangherlini black holes as its solutions in Sec. \ref{sec6}. (vi) Finally, concluding remarks and discussions  will be given in Sec. \ref{con}.  
\section{Four-dimensional ghost-free nonlinear massive gravity} \label{sec2}
In this section, we will remark some basic details of the four-dimensional ghost-free nonlinear massive gravity, which have already been investigated extensively.  As a result, they will be useful for comparing the nonlinear massive gravity with its higher dimensional scenarios.  An action of four-dimensional ghost-free nonlinear massive gravity proposed by de Rham, Gabadadze, and Tolley  ~\cite{RGT} is given by 
\begin{equation} \label{action}
S = \frac{M_p^2}{2}\int {d^4 x \sqrt {-g}} \Bigl\{ {R +m_g^2\left({{\cal L}_2+\alpha_3 {\cal L}_3+\alpha_4{\cal L}_4}\right)} \Bigr\},
\end{equation}
where $M_p$ is the Planck mass,  $m_g$ is the graviton mass, $\alpha_{3,4}$ are free parameters, and the massive terms ${\cal L}_i$ ($i=2-4$) are defined as
\begin{eqnarray}
 \label{eq1}
{\cal L}_2 &=& [{\cal K}]^2- [{\cal K}^2], \\
 \label{eq2}
{\cal L}_3 &=  &  \frac{1}{3}  [{\cal K}]^3-[{\cal K}] [{\cal K}^2]+\frac{2}{3}[{\cal K}^3] , \\
 \label{eq3}
{\cal L}_4 &=&   \frac{1}{12}[{\cal K}]^4-\frac{1}{2}[{\cal K}]^2 [{\cal K}^2]+\frac{1}{4} [{\cal K}^2]^2+\frac{2}{3}[{\cal K}][{\cal K}^3]-\frac{1}{2}[{\cal K}^4].\nonumber\\
\end{eqnarray}
Note that the above square brackets are defined  as~\cite{RGT,review,NAH}
\begin{equation} \label{eq4}
 [{\cal K}] \equiv \text{tr}{\cal K}^\mu{ }_{\nu}; [{\cal K}]^2 \equiv \left({\text{tr}{\cal K}^\mu{ }_\nu}\right)^2;  [{\cal K}^2] \equiv \text{tr}{\cal K}^\mu{ }_\alpha {\cal K}^\alpha{ }_\nu,
\end{equation}
where 
\begin{eqnarray}
 \label{eq5}
{\cal K}^\mu{ }_\nu &= & \delta^\mu{ }_\nu - M^{\mu}{ }_{\nu},  \\
Z_{\mu\nu}&=& f_{ab}\partial_\mu \phi^a \partial_\nu \phi^b, \\
\label{eq6}
 Z^{\mu}{ }_{\nu}&\equiv &  g^{\mu\alpha}Z_{\alpha\nu}\equiv M^{\mu}{ }_{\rho}M^{\rho}{ }_{\nu}.
\end{eqnarray}
Here, according to  Ref. ~\cite{RGT}, $g_{\mu\nu}$ is the physical metric, $f_{ab}$ is the fiducial (or reference) metric, and $\phi^a$ ($a=0-3$) are the St\"uckelberg scalar fields introduced to give a manifestly diffeomorphism invariant description ~\cite{NAH}.  In the massive gravity framework, the fiducial metric has been assumed to be non-dynamical, i.e.,  time derivatives of scale factors of the fiducial metric no longer appear in the massive gravity Lagrangian. Hence, constraint equations associated with the fiducial metric become algebraic equations rather than differential equations ~\cite{WFK}. And by making the fiducial metric dynamical, i.e., the fiducial metric plays similarly as the physical metric does, we will obtain the so-called bigravity ~\cite{SFH}.  Note again that the dRGT theory with an arbitrary fiducial metric has been confirmed to be free of the BD ghost through  different approaches ~\cite{proof}.

As a result,  varying the action (\ref{action}) with respect to the physical metric  leads to the modified Einstein field equation as ~\cite{WFK}
\begin{equation} \label{Einstein}
\left({R_{\mu\nu}-\frac{1}{2}Rg_{\mu\nu}}\right)+m_g^2 \left({ X_{\mu\nu}+\alpha_4 Y_{\mu\nu}}\right)=0,
\end{equation}
with
\begin{eqnarray} \label{eqX}
X_{\mu\nu}&=&  -\frac{1}{2} \left(\alpha {\cal L}_2 +\beta {\cal L}_3 \right) g_{\mu\nu} + \tilde X_{\mu\nu},\\
\tilde X_{\mu\nu}&=& {\cal K}_{\mu\nu} -[{\cal K}]g_{\mu\nu} -\alpha \left\{{{\cal K}_{\mu\nu}^2-[{\cal K}]{\cal K}_{\mu\nu} }\right\} \nonumber\\
&& +\beta \left\{{{\cal K}_{\mu\nu}^3-[{\cal K}]{\cal K}_{\mu\nu}^2+\frac{{\cal L}_2}{2} {\cal K}_{\mu\nu} }\right\}, 
\end{eqnarray}
\begin{eqnarray}
\label{eqY}
 Y_{\mu\nu} &=& -\frac{{\cal L}_4}{2} g_{\mu\nu} + \tilde Y_{\mu\nu},\\
 \label{eq-hatY}
\tilde Y_{\mu\nu} &=& \frac{{\cal L}_3}{2} {\cal K}_{\mu\nu}  -\frac{{\cal L}_2}{2}  {\cal K}^2_{\mu\nu} +[{\cal K}]{\cal K}^3_{\mu\nu} -{\cal K}^4_{\mu\nu}.
\end{eqnarray}
Here,  $\alpha = \alpha_3+1$, $\beta =\alpha_3+\alpha_4$, ${\cal K}_{\mu\nu}=g_{\mu\alpha_1}{\cal K}^{\alpha_1}{ }_\nu$ and ${\cal K}_{\mu\nu}^n=g_{\mu\alpha_1}{\cal K}^{\alpha_1}{ }_{\alpha_2}... {\cal K}^{\alpha_n}{ }_\nu$ ($n \geq 2$). As will be shown later, $Y_{\mu\nu}$ can be proved to be zero in general as a consequence of the Cayley-Hamilton theorem, which is associated with the definition of matrix determinant. Note that the Einstein field equations without the expression of $Y_{\mu\nu}$ have also been derived in Refs. ~\cite{RGT1,RGT2}. 

In addition, according to Ref. ~\cite{WFK} the modified Einstein equations (\ref{Einstein}) can be rewritten as
\begin{equation}
\left({R_{\mu\nu}-\frac{1}{2}R g_{\mu\nu}}\right)-\frac{m_g^2}{2}{\cal L}_M^0 g_{\mu\nu} = 0
\end{equation}
along with the following constraint equations associated with the fiducial metric:
\begin{equation}
t_{\mu\nu}\equiv \tilde X_{\mu\nu}+\alpha_4  \tilde Y_{\mu\nu}-\frac{1}{2}\left(\alpha_3 {\cal L}_2+\alpha_4 {\cal L}_3 \right)g_{\mu\nu}=0,
\end{equation}
where ${\cal L}_M^0 \equiv {\cal L}_2 + \alpha_3 {\cal L}_3 + \alpha_4 {\cal L}_4 $. More interestingly, the massive graviton terms can be shown to be an effective cosmological constant for a large class of metric spaces, provided that the fiducial metric is compatible with the physical metric ~\cite{WFK}. In particular, given both diagonal Bianchi type I physical and fiducial metrics, it has been shown in Ref. ~\cite{WFK} that
\begin{equation}
m_g^2 {\cal L}_M^0 = -2 \Lambda_M^0,
\end{equation}
where $\Lambda_M^0$, an effective cosmological constant, is defined as follows
\begin{eqnarray} \label{lambda1}
\Lambda_M^0 &=&\frac{3 m_g^2}{2\alpha_4^3} \biggl[9 \alpha_3^4 +6 \alpha_4^2 -18\alpha_3^2 \alpha_4  \nonumber\\
&&  \pm \alpha_3 \left(3\alpha_3^2-4\alpha_4\right) \sqrt{3\left(3\alpha_3^2-4\alpha_4\right) }\biggr]
\end{eqnarray}
for $\alpha_3^2>4\alpha_4/3$ and 
\begin{equation}\label{lambda2}
\Lambda_M^0 =\frac{m_g^2}{\alpha_4 -\alpha_3^2}
\end{equation}
for $\alpha_3^2<4\alpha_4/3$.

In addition, both $\Lambda_M^0$'s shown above will become $-4m_g^2/\alpha_3^2<0$ when  $\alpha_4 = 3\alpha_3^2/4 >0 $. Furthermore, the Bianchi type I expanding solutions of the four-dimensional dRGT theory have been shown to be stable against field perturbations for both isotropic FLRW fiducial  ~\cite{bianchi-I} and  Bianchi type I fiducial metrics ~\cite{WFK}. Hence, the so-called cosmic no-hair conjecture proposed by Hawking and his colleagues ~\cite{Hawking,WFK1}, which states that the final state of universe should be isotropic, seems to be violated in the context of dRGT theory. 

Since the Bianchi type I metric is a generalization of FLRW metric, it is straightforward to show that in the isotropic FLRW limit we can obtain the following effective cosmological constant as shown in Eq. (\ref{lambda1}) ~\cite{WFK}. Consequently, the dRGT theory with the positive $\Lambda_M^0$ defined in Eq. (\ref{lambda1})  can be shown to admit the de Sitter metric:
\begin{equation}
ds_{dS}^2 =-dt^2 +\exp\left[2\sqrt{\frac{\Lambda_M^0}{3}}~t\right]\left(dx^2+dy^2+dz^2\right),
\end{equation}
 as its cosmological solution if both physical and fiducial metrics are taken to be FLRW one. 
More interestingly, we will show below that  the values of the effective cosmological constant shown in Eq. (\ref{lambda1}) will also be recovered  for some metrics of five-dimensional  nonlinear massive gravity under an assumption that both physical and fiducial metrics are compatible with each other. 
\section{Higher dimensional nonlinear massive gravity} \label{sec3}
\subsection{Cayley-Hamilton theorem and ghost-free graviton terms}
First, we will present a close relation between the Cayley-Hamilton theorem and the strategy to construct the ghost-free graviton terms ${\cal L}_i$ ($i=2-4$) ~\cite{RGT}.   Before going to discuss in details, we would like to mention that there have been some consistent analysis on re-constructing the four-dimensional massive graviton terms of dRGT theory, e.g., Refs. ~\cite{TMN,SFH}, especially Ref. ~\cite{Bonifacio:2014rba}. We hope that  our present analysis  together with them could shed more light on the mathematical structure of ghost-free nonlinear massive gravity in four and higher dimensional spaces. 

In mathematics, e.g., see Ref. ~\cite{CH-theorem}, there exists the well-known Cayley-Hamilton theorem, which states that  any square matrix must obey its characteristic equation. In particular, given a $n\times n$ matrix $K$ with its characteristic equation, ${\cal P}(\lambda) \equiv \det(\lambda I_n- K)=0$, then 
\begin{eqnarray} \label{characteristic-equation}
{\cal P}(K) &\equiv& K^n- {\cal D}_{n-1}K^{n-1}+{\cal D}_{n-2}K^{n-2}-... \nonumber\\
&&+ (-1)^{n-1} {\cal D}_1  K+ (-1)^{n} \det(K) I_n =0, \nonumber\\
\end{eqnarray}
where ${\cal D}_{n-1} = \text{tr} K \equiv [K]$ and ${\cal D}_{n-j}$ ($2\leq j\leq n-1$) are coefficients of the characteristic polynomial. In addition, $I_n$ is a $n\times n$ identity matrix.

In particular, for $n=2$ we have the following characteristic equation as
\begin{equation}
K^2 - [K] K + \det K_{2\times 2} I_2=0,
\end{equation}
which implies 
\begin{equation}
\det K_{2\times 2} = \frac{1}{2} \Bigl\{ [K]^2 -[K^2] \Bigr\},
\end{equation}
after taking the trace.  It is clear that the definition of $\det K_{2\times 2}$ looks similar to that of four-dimensional graviton term ${\cal L}_2/2$.

On the other hand, for $n=3$ the corresponding characteristic equation turns out to be
\begin{equation}
K^3 - [K] K^2 + \frac{1}{2} \left\{ [K]^2 -[K^2] \right\} K + \det K_{3\times 3} I_3 =0, 
\end{equation} 
which leads to
\begin{equation}
\det K_{3 \times 3} =\frac{1}{6} \Bigl\{ [K]^3 -3 [K^2] [K] +2[K^3] \Bigr\}.
\end{equation}
It is clear that the definition of $\det K_{3\times 3}$ looks similar to that of  four-dimensional graviton term ${\cal L}_3/2$.

Similarly, for $n=4$ the corresponding characteristic equation is given by
\begin{eqnarray} \label{characteristic-n=4}
&& K^4 -[K] K^3 +  \frac{1}{2} \left\{ [K]^2 -[K^2] \right\} K^2 \nonumber\\
&& - \frac{1}{6} \left\{ [K]^3 -3 [K^2] [K] +2[K^3] \right\} K +\det K_{4\times4} I_4 =0,\nonumber\\
\end{eqnarray}
which gives
\begin{eqnarray}
\det K_{4\times4} &=& \frac{1}{24} \Bigl\{ [K]^4 -6[K]^2 [K^2] +3[K^2]^2  \nonumber\\
&&  +8[K][K^3] -6[K^4]\Bigr\}.
\end{eqnarray}
It is straightforward to see that if use ${\cal K}$ to denote $K_{4\times4}$ then
\begin{equation}
\det K_{4\times4} =\det {\cal K}=\frac{{ \cal L}_4}{2},
\end{equation}
where ${ \cal L}_4$ has been defined in Eq. (\ref{eq3}).  More interestingly, we will be able to show  $Y_{\mu\nu}$ defined in Eq. (\ref{eqY}) always vanishes as a consequence of the Cayley-Hamilton theorem. Indeed, we rewrite $\tilde Y_{\mu\nu}$ defined in Eq. (\ref{eq-hatY}) as
\begin{equation}
\tilde Y_{\mu\nu}=g_{\mu\alpha} {\cal Q}^{\alpha}{ }_{\nu},
\end{equation}
with
\begin{eqnarray}
{\cal Q}^{\alpha}{ }_{\nu} &=& - \left({\cal K}^4\right)^{\alpha}{ }_{\nu}+ [{\cal K}]\left({\cal K}^3\right)^{\alpha}{ }_{\nu} -\frac{1}{2}\left({[{\cal K}]^2-[{\cal K}^2] }\right)\left({\cal K}^2\right)^{\alpha}{ }_{\nu} \nonumber\\
&& +\frac{1}{6}\left({ [{\cal K}]^3-3[{\cal K}] [{\cal K}^2]+2 [{\cal K}^3] }\right) {\cal K}^{\alpha}{ }_{\nu}.
\end{eqnarray}
And according to the characteristic equation in four dimensions (\ref{characteristic-n=4}), it is now clear that
\begin{equation}
Y_{\mu\nu} = g_{\mu\alpha}\left[ {\cal Q}^{\alpha}{ }_{\nu}- \frac{{ \cal L}_4}{2} \delta^{\alpha}{ }_{\nu} \right] =0,
\end{equation}
as claimed.  

Next, we will show the ghost-free property of dRGT graviton terms  is indeed a consequence of the Cayley-Hamilton theorem. Recall the tensor $X^{(4)}_{\mu\nu}$ defined in  Ref. ~\cite{RGT}:
\begin{equation}
X^{(n)}_{\mu\nu} (g_{\mu\nu}, {\cal K}) = \sum\limits_{m=0}^n (-1)^m\frac{n!}{2(n-m)!}{\cal K}^{m}_{\mu\nu}{\cal L}_{\text{der}}^{(n-m)}({\cal K}),
\end{equation}
with replacements, $\eta_{\mu\nu} \to g_{\mu\nu}$, $\Pi \to {\cal K}$, and ${\cal K}^{0}_{\mu\nu}=g_{\mu\nu}$.  Note that ones claimed   in Ref.~\cite{RGT} that $X^{(4)}_{\mu\nu}(\eta_{\mu\nu}, \Pi)=0$.  Similarly, we can also be able to show that $X^{(4)}_{\mu\nu}(g_{\mu\nu}, {\cal K})=0$. As a result, $X^{(4)}_{\mu\nu}(g_{\mu\nu}, {\cal K})$ is given by
\begin{eqnarray}
X^{(4)}_{\mu\nu} &=& \frac{{ \cal L}^{(4)}_{\text{der}}}{2} g_{\mu\nu} -2 {\cal K}_{\mu\nu} { \cal L}^{(3)}_{\text{der}} +6{\cal K}^2_{\mu\nu}{ \cal L}^{(2)}_{\text{der}}-12 {\cal K}^3_{\mu\nu}{ \cal L}^{(1)}_{\text{der}} \nonumber\\
&& + 12 {\cal K}^4_{\mu\nu}{ \cal L}^{(0)}_{\text{der}},
\end{eqnarray}
where $ { \cal L}^{(n)}_{\text{der}}$  is defined in Ref. ~\cite{RGT} as
\begin{equation}
 { \cal L}^{(n)}_{\text{der}}({\cal K}) = -\sum\limits_{m=1}^n (-1)^m\frac{(n-1)!}{(n-m)!}[{\cal K}^{m}]{\cal L}_{\text{der}}^{(n-m)}({\cal K}),
\end{equation}
with $ { \cal L}^{(0)}_{\text{der}}=1$ and $\Pi$ has been replaced by ${\cal K}$. The explicit formulas of $ { \cal L}^{(n)}_{\text{der}}$ with $n=1-4$ are therefore given by
\begin{eqnarray}
 { \cal L}^{(1)}_{\text{der}} &=& [{\cal K}], \\
 { \cal L}^{(2)}_{\text{der}} &= &[{\cal K}]^2 -[{\cal K}^2], \\
 { \cal L}^{(3)}_{\text{der}} &=& [{\cal K}]^3-3[{\cal K}] [{\cal K}^2]+2 [{\cal K}^3], \\
 { \cal L}^{(4)}_{\text{der}} &=& [{\cal K}]^4-6[{\cal K}]^2 [{\cal K}^2]+3 [{\cal K}^2]^2+8 [{\cal K}][{\cal K}^3]-6[{\cal K}^4] ,\nonumber\\
\end{eqnarray}
Thanks to these definitions, we finally arrive at 
\begin{equation}
X^{(4)}_{\mu\nu} = 12 \left({\frac{{ \cal L}_4}{2} g_{\mu\nu} - \tilde Y_{\mu\nu} }\right) = - 12 Y_{\mu\nu}.
\end{equation}
Since $Y_{\mu\nu}=0$  then $X^{(4)}_{\mu\nu}=0$  as expected. Note that there is the recursive relation ~\cite{RGT}  as 
\begin{equation}
 X^{(n)}_{\mu\nu}  = -n {\cal K}^\alpha_\mu  X^{(n-1)}_{\alpha\nu} + {\cal K}^{\alpha\beta} X^{(n-1)}_{\alpha\beta} g_{\mu\nu}.
\end{equation} 
Hence  $X^{(n>4)}_{\mu\nu} = 0$ since $ X^{(n=4)}_{\mu\nu} =0$ as shown above.  This is a guarantee that no ghost-like pathology arise at the quartic or higher order levels with arbitrary physical and fiducial metrics, consistent with  investigations in Ref. ~\cite{RGT}. Note that the result $X^{(n \geq 4)}_{\mu\nu} = 0$ is equivalent to the result that all graviton terms ${\cal L}_{n \geq 5}$  vanish identically in four-dimensional spacetime ~\cite{WFK}.

It is now clear that the massive graviton terms are closely related to the Cayley-Hamilton theorem. In other words, the Cayley-Hamilton theorem really shows us the useful way  to construct the graviton terms, which have been shown to be free of the BD ghost ~\cite{RGT}. 

Thanks to this useful method,  we are able to construct the graviton terms in $n \geq 5$ spacetimes, which ensure that the following higher dimensional nonlinear massive gravity  is ghost-free. For a heuristic reason, we will show the explicit definition of graviton term, ${\cal L}_5$,  in five-dimensional spacetime as a specific example. Indeed, for $n=5$ we have the following characteristic equation:
\begin{eqnarray}
&&K^5 - [K]K^4 +  \frac{1}{2} \left\{ [K]^2 -[K^2] \right\} K^3\nonumber\\
&& - \frac{1}{6} \left\{ [K]^3 -3 [K^2] [K] +2[K^3] \right\} K^2 \nonumber\\
&&+ \frac{1}{24} \left\{ [K]^4 -6[K]^2[K^2] +3[K^2]^2 +8[K][K^3] -6[K^4]\right\} K\nonumber\\
&& - \det K_{5\times 5} I_5 =0.
\end{eqnarray}
By taking the trace, we arrive at
\begin{eqnarray} \label{detK-5d}
\frac{{\cal L}_5}{2}&= &\det K_{5\times 5} \nonumber\\
&=& \frac{1}{120} \Bigl\{ [K]^5 -10 [K]^3 [K^2] +20[K]^2 [K^3]  \nonumber\\
&&    -20  [K^2][K^3]+15[K]  [K^2]^2 -30[K] [K^4]   \nonumber\\
&&  +24 [K^5] \Bigr\}. 
\end{eqnarray}
This expression of ${\cal L}_5$ looks similar to that of ${ \cal L}^{(5)}_{\text{der}}(\Pi)$ defined by Eq. (31) in the second paper of Ref. ~\cite{RGT}. Note again that ${\cal L}_5$ (or ${ \cal L}^{(5)}_{\text{der}}$) must vanish in any four-dimensional spacetime. However,  this result will no longer hold in higher-than-four dimensional spacetimes.

It is straightforward to show that $X^{(n\geq 5)}_{\mu\nu}=0$ in the five-dimensional spacetime, which ensures that any ghost-like pathology arising at the quintic or higher order levels must disappear, no matter the form of physical and fiducial metrics. By doing the same steps, we can define the other graviton terms in spacetimes, whose number of dimension is larger than five. Indeed,  we will present here  graviton terms in six- and seven-dimensional spacetimes as specific demonstrations for our claim. Given the definition of determinant of matrices $K_{n\times n}$ with $n=2,~3,~4$, and $5$, we are able to define the determinant of six- and seven-dimensional matrices, $K_{6\times 6}$  and $K_{7\times 7}$, to be
\begin{widetext}
\begin{eqnarray}\label{L6}
\frac{{\cal L}_6}{2}&=& \det K_{6\times 6} \nonumber\\
&=& \frac{1}{720} \Bigl\{ [K]^6 -15[K]^4 [K^2]+40[K]^3 [K^3] - 90 [K]^2 [K^4]+45 [K]^2 [K^2]^2  -15 [K^2]^3    \nonumber\\
&&  +40 [K^3]^2 -120  [K^3] [K^2] [K]+90[K^4] [K^2]   +144 [K^5] [K]-120 [K^6] \Bigr\},\\
\label{L7}
\frac{{\cal L}_7}{2} &=& \det K_{7\times 7} \nonumber\\
&=&\frac{1}{5040} \Bigl\{[K]^7 -21 [K]^5[K^2]+70 [K]^4[K^3] -210[K]^3 [K^4]+105[K]^3[K^2]^2 -420[K]^2[K^2][K^3] \nonumber\\
&& +504[K]^2[K^5]-105[K^2]^3[K]+210[K^2]^2 [K^3]-504[K^2][K^5] +280[K^3]^2[K] -420[K^3] [K^4]  \nonumber\\
&&+630[K^4] [K^2] [K]-840[K^6][K]+720[K^7] \Bigr\},
\end{eqnarray}
\end{widetext}
respectively. 
Similarly,  it turns out that any ghost-like pathology arising at the sixth or higher order levels will no longer exist due to the fact that  $X^{(n\geq 6)}_{\mu\nu}=0$ in the six-dimensional spacetime. This result can also be extended for seven-dimensional spacetime, i.e., any ghost-like pathology arising at the seventh or higher order levels  will disappear due to the fact that  $X^{(n\geq 7)}_{\mu\nu}=0$. Before going to the next subsection, we would like to note that the expressions of ${\cal L}_5$ and ${\cal L}_6$ have been mentioned in Ref.~\cite{Bonifacio:2014rba}.
\subsection{Five-dimensional nonlinear massive gravity}
Generally, we can conclude that all graviton terms ${\cal L}_{n+i}$ with $i=1,2,3,...$ must disappear in the $n$-dimensional spacetime, according to the Cayley-Hamilton theorem. In other words, using the characteristic equation (\ref{characteristic-equation}) we can construct BD ghost-free graviton terms for $n$-dimensional nonlinear massive gravity theory of  ${\cal K}_{n\times n}$ matrix.  For heuristic reasons,  we will discuss a five-dimensional nonlinear massive gravity and its cosmological implications in the rest of the present paper. 

As shown above, an action of five-dimensional ghost-free nonlinear massive gravity is given by 
\begin{eqnarray} 
S &=& \frac{M_p^2}{2}\int {d^5 x \sqrt {-g}} \nonumber\\
&& \times \Bigl\{ {R +m_g^2\left({{\cal L}_2+\alpha_3 {\cal L}_3 +\alpha_4 {\cal L}_4 +\alpha_5 {\cal L}_5} \right)} \Bigr\},\nonumber\\
\end{eqnarray}
here  $\alpha_{5}$ is an additional field parameter associated with the massive terms ${\cal L}_5$ defined as
\begin{eqnarray}
{\cal L}_5 &=& \frac{1}{60}[{\cal K}]^5 -\frac{1}{6}[{\cal K}]^3[{\cal K}^2]+\frac{1}{3}[{\cal K}]^2  [{\cal K}^3] -\frac{1}{3}[{\cal K}^2] [{\cal K}^3] \nonumber\\
&&+\frac{1}{4} [{\cal K}] [{\cal K}^2]^2   -\frac{1}{2}[{\cal K}][{\cal K}^4] +\frac{2}{5}[{\cal K}^5]. 
\end{eqnarray}
Here, we have used the result shown in Eq. (\ref{detK-5d}) with $K={\cal K}$. Additionally, note again that all terms ${\cal L}_n$ with $n\geq 6$ will vanish identically in the five-dimensional spacetime.

As a result, the corresponding five-dimensional Einstein field equations turn out to be
\begin{equation} \label{Einstein-5d}
\left({R_{\mu\nu}-\frac{1}{2}Rg_{\mu\nu}}\right)+m_g^2 \left({ { X}_{\mu\nu}+ \sigma {Y}_{\mu\nu} +\alpha_5 {W}_{\mu\nu}}\right)=0,
\end{equation}
where 
\begin{eqnarray} \label{eqX-5d}
X_{\mu\nu}&=&  -\frac{1}{2} \left(\alpha {\cal L}_2 +\beta {\cal L}_3 \right) g_{\mu\nu} + \tilde X_{\mu\nu},\\
\tilde X_{\mu\nu}&=& {\cal K}_{\mu\nu} -[{\cal K}]g_{\mu\nu} -\alpha \left\{{{\cal K}_{\mu\nu}^2-[{\cal K}]{\cal K}_{\mu\nu} }\right\} \nonumber\\
&& +\beta \left\{{{\cal K}_{\mu\nu}^3-[{\cal K}]{\cal K}_{\mu\nu}^2+\frac{{\cal L}_2}{2} {\cal K}_{\mu\nu} }\right\}, 
\end{eqnarray}
\begin{eqnarray}
\label{eqY-5d}
 Y_{\mu\nu} &=& -\frac{{\cal L}_4}{2} g_{\mu\nu} + \tilde Y_{\mu\nu},\\
\tilde Y_{\mu\nu} &=& \frac{{\cal L}_3}{2} {\cal K}_{\mu\nu}  -\frac{{\cal L}_2}{2}  {\cal K}^2_{\mu\nu} +[{\cal K}]{\cal K}^3_{\mu\nu} -{\cal K}^4_{\mu\nu},
\end{eqnarray}
\begin{eqnarray}
W_{\mu\nu} &=& -\frac{{\cal L}_5}{2}g_{\mu\nu} + \tilde W_{\mu\nu}, \\
\tilde W_{\mu\nu} &=& \frac{{\cal L}_4}{2}  {\cal K}_{\mu\nu} -\frac{{\cal L}_3}{2} {\cal K}^2_{\mu\nu}+\frac{{\cal L}_2}{2} {\cal K}^3_{\mu\nu} - [{\cal K}]{\cal K}^4_{\mu\nu} +{\cal K}^5_{\mu\nu}, \nonumber\\
\end{eqnarray}
with $\alpha = \alpha_3+1$, $\beta =\alpha_3+\alpha_4$, and $\sigma =\alpha_4+\alpha_5$. Note that although the tensors $X_{\mu\nu}$ and $Y_{\mu\nu}$ shown in Eqs. (\ref{eqX-5d}) and (\ref{eqY-5d})  look similar to that  in Eqs. (\ref{eqX}) and (\ref{eqY}) but they now live in five-dimensional spacetime. Hence, they will contain more terms than that defined in four-dimensional spacetime

It is noted that in the four-dimensional spacetime, where $W_{\mu\nu}$ no longer exists, the tensor $Y_{\mu\nu}$ has been shown to be zero as a consequence of the Cayley-Hamilton theorem. And in five-dimensional spacetime, we also have the same result for the tensor $W_{\mu\nu}$, i.e., $W_{\mu\nu}=0$ in general as a consequence of the Cayley-Hamilton theorem. Note again that $Y_{\mu\nu}\neq 0$ in  higher-than-four dimensional spacetimes. Of course, $W_{\mu\nu}\neq 0$ in higher-than-five dimensional spacetimes.

Following detailed analysis of Ref. ~\cite{WFK}, the modified Einstein field equations (\ref{Einstein-5d}) can be rewritten as 
\begin{equation} \label{Einstein-5d-reduced}
\left({R_{\mu\nu}-\frac{1}{2}Rg_{\mu\nu}}\right) -\frac{m_g^2}{2}{\cal L}_M g_{\mu\nu}=0,
\end{equation}
along with the corresponding constraint equations associated with the existence of fiducial metric:
\begin{eqnarray}\label{5d-constraints}
t_{\mu\nu}&\equiv& \tilde X_{\mu\nu}+\sigma \tilde Y_{\mu\nu}+\alpha_5\tilde W_{\mu\nu}\nonumber\\
&& -\frac{1}{2} \left(\alpha_3 {\cal L}_2+\alpha_4 {\cal L}_3+ \alpha_5 {\cal L}_4 \right)g_{\mu\nu} = 0.
\end{eqnarray}
Here ${\cal L}_M$ is given by
\begin{equation}
{\cal L}_M = {\cal L}_2+\alpha_3 {\cal L}_3 +\alpha_4 {\cal L}_4 +\alpha_5 {\cal L}_5.
\end{equation} 
The equation (\ref{Einstein-5d-reduced}) implies an important result that the massive graviton terms act as an effective cosmological constant $\Lambda_M \equiv - {m_g^2 {\cal L}_M}/2$ due to the consequence of the Bianchi identity: $\partial^\nu {\cal L}_M=0$. This result can be applied for a large class of physical and fiducial metrics. Indeed, we will show below the constant-like behavior of graviton terms for some well-known metrics under an assumption that the fiducial metric is compatible with the physical one. Note that the constraint equations (\ref{5d-constraints}) associated with the fiducial metric do not involve the Einstein tensor, then they are indeed algebraic equations rather than differential equations if  the  St\"uckelberg scalar fields are taken to be in the unitary gauge, i.e., $\phi^a =x^a$. 
\section{Five-dimensional FLRW metrics} \label{sec4}
In five-dimensional spacetime, we will consider the FLRW physical and fiducial metrics  given by 
\begin{eqnarray} \label{frw-1}
ds_{5d}^2 (g_{\mu\nu}) &=& -N_1^2(t)dt^2 + a_1^2(t) \left(d\vec{x}^2+du^2 \right), \\
 \label{frw-2}
Z^{5d}_{\mu\nu}(f_{ab})   &= &  -N_2^2(\phi^0)\partial_\mu\phi^0\partial_\nu\phi^0+a_2^2(\phi^0) \sum\limits_{a=1}^4 \partial_\mu\phi^a \partial_\nu\phi^a , \nonumber\\
\end{eqnarray}
where $a_i$'s ($i=1-2$) are scale factors and $u$ is the fifth dimension ~\cite{5d-FRW}.  In addition, $N_1$ and $N_2$ are the lapse functions, which are introduced to obtain the following Friedmann equations from their Euler-Lagrange equations ~\cite{RGT,Arnowitt:1962hi,WFK}. Note that we can set $N_1=1$ after  its corresponding Friedmann equation is derived. However, we should not do the same thing for $N_2$, i.e., $N_2$ should be regarded as a free field variable.  Note again  we have assumed that the fiducial metric is compatible with the physical one in the present paper. Following Refs.~\cite{KNT,RGT2,TMN,RGT1,bianchi-I,WFK}, the  St\"uckelberg scalar fields will be taken to be in the unitary gauge, $\phi^a =x^a$, such that $\partial_\mu \phi^a =\delta_\mu^a$. It then will follow that
\begin{equation} \label{frw-3}
ds_{5d}^2 (Z_{\mu\nu}) = -N_2^2(t)dt^2 + a_2^2(t) \left(d\vec{x}^2 +du^2 \right).
\end{equation}
It will be convenient by defining the following expression:
\begin{eqnarray} \label{frw-4}
\left[{\cal K} \right]^n &=& \left(5-\gamma-4\Sigma \right)^n, \nonumber\\
\left[{\cal K}^n \right] &=& \left(1-\gamma \right)^n +4\left(1-\Sigma\right)^n  , \nonumber \\
\gamma &=& \frac{N_2}{N_1}; ~ \Sigma =\frac{a_2}{a_1}.
\end{eqnarray}
Hence, the corresponding graviton terms turn out to be
\begin{eqnarray} \label{frw-5}
{\cal L}_2 &=&2 \left[3 \Sigma^2  +3\left(\gamma-3\right)\Sigma +3\left(2-\gamma\right) \right] \nonumber\\
&&+ 2\left(3\Sigma+\gamma-4\right)\left(\Sigma-1\right),\\
 \label{frw-6}
{\cal L}_3 &=& -2 \left[\Sigma^3   +3 \left(\gamma-2 \right) \Sigma^2+ 3 \left(3-2\gamma \right) \Sigma+3\gamma-4\right] \nonumber\\
&& -2  \left[ 3 \Sigma^2 + 3 \left(\gamma-3\right)\Sigma +3 \left(2-\gamma \right)\right]\left(\Sigma-1\right), \\
 \label{frw-7}
{\cal L}_4&=& 2 \left(\gamma-1\right) \left(\Sigma-1 \right)^3 +2  \left[ \Sigma^3  + 3 \left(\gamma-2 \right) \Sigma^2 \right. \nonumber\\
&& \left. +3 \left(3-2\gamma \right) \Sigma+3\gamma-4\right] \left(\Sigma-1\right), \\
 \label{frw-8}
{\cal L}_5&=&-2 \left(\gamma-1 \right) \left(\Sigma-1\right)^4 .
\end{eqnarray}
Furthermore, taking a summation of all graviton terms ${\cal L}_i$ ($i=2-5$) leads to
\begin{eqnarray}  \label{frw-9}
{\cal L}_M &=& 2 \Bigl[{\left({\gamma\alpha_4-\gamma_3}\right)\Sigma^3
+3\left({\gamma_2-\gamma\gamma_3}\right)\Sigma^2 } \nonumber\\
&& { +3 \left({\gamma\gamma_2-\gamma_1}\right) \Sigma -\gamma \gamma_1 +\left({3\gamma_1-3\gamma_2+\gamma_3}\right) } \Bigr]  \nonumber\\
&& + 2 \Bigl\{\alpha_4 \Sigma^3 -\alpha_5 \left(\gamma-1\right) \left(\Sigma-1\right)^3   \nonumber\\
&& -3\left[\gamma_3- \left(\gamma-1\right)\alpha_4 \right] \Sigma^2  \nonumber\\
&& +  3 \left[\gamma_2 -\left(\gamma -1\right)  \left(\gamma_3+\alpha_4 \right) \right] \Sigma   \nonumber\\
&&   + \left(\gamma-1\right) \left(3\gamma_3 +1 \right) -\gamma_1 \Bigr\} \left(\Sigma-1\right),
\end{eqnarray}
with the parameters $\gamma_i$'s defined as ~\cite{bianchi-I}
\begin{equation} \label{def-gamma}
\gamma_1 = 3+3\alpha_3+\alpha_4; ~\gamma_2=1+2\alpha_3+\alpha_4;~ \gamma_3=\alpha_3+\alpha_4.
\end{equation}

Following the method used in Ref. ~\cite{WFK}, we will  solve constraint equations associated with the scale factors of fiducial metric in order to show the constant-like behavior of graviton terms. In particular, these constraint equations are the Euler-Lagrange equations of $N_2$ and $a_2$:
\begin{equation} \label{frw-Euler}
\frac{\partial {\cal L}_M}{\partial N_2}=0; ~ \frac{\partial {\cal L}_M}{\partial a_2}=0.
\end{equation}
Here, we note that there is no any time derivative of $N_2$ and $a_2$ in the massive graviton Lagrangian ${\cal L}_M$. As a result, the above constraint equations are equivalent with the following equations:
\begin{equation} \label{frw-Euler-1}
\frac{\partial {\cal L}_M}{\partial \gamma}=0; ~ \frac{\partial {\cal L}_M}{\partial \Sigma}=0,
\end{equation}
since $\gamma$ and $\Sigma$ have been set as functions of $N_2$ and $a_2$, respectively.  Thanks to the explicit definition of ${\cal L}_M$ as shown in Eq. (\ref{frw-9}), the constraint equations (\ref{frw-Euler-1}) can be expanded to be
\begin{eqnarray}\label{frw-10}
 \alpha_5 \hat\Sigma^3+4\alpha_4 \hat\Sigma^2+6\alpha_3 \hat\Sigma+4 &=&0, \\
\label{frw-11}
 \left(\alpha_5 \hat\Sigma^3 +3\alpha_4 \hat\Sigma^2 +3\alpha_3 \hat\Sigma +1\right)\hat\gamma && \nonumber\\
+ \left(\alpha_4 \hat\Sigma^2+3\alpha_3 \hat\Sigma+3\right)\hat\Sigma &=&0,
\end{eqnarray}
where $\hat\Sigma \equiv 1-\Sigma$ and $\hat\gamma \equiv 1-\gamma$ as additional variables. Note that these equations can also be obtained from  the (tensor) constraint equations defined in Eq. (\ref{5d-constraints}). In this paper,  however, we prefer using  the Euler-Lagrange equations since they turn out to be more effective and convenient than the tensor equations (\ref{5d-constraints}).  It is straightforward to show that Eq. (\ref{frw-11}) can be solved, with the help of Eq. (\ref{frw-10}), to give a solution:
\begin{equation}
\hat\gamma= \hat\Sigma,
\end{equation}
assuming that $\alpha_4 \hat\Sigma^2+3\alpha_3 \hat\Sigma+3\neq 0$. As a result, the cubic equation  (\ref{frw-10}) of $\hat\Sigma$ can be solved to admit three real or complex solutions by the standard methods, e.g., the Cardano's method. In order to investigate whether the above cubic equation admits complex solution(s), we define the following discriminant:
\begin{eqnarray} \label{delta}
\Delta = 16 \left[108\alpha_3 \alpha_4 \alpha_5 -16 \alpha_4^3 +36\alpha_3^2 \alpha_4^2 - 9\alpha_3^2 \alpha_5 -27 \alpha_5^2\right]. \nonumber\\
\end{eqnarray}
It is known that if $\Delta <0$ the cubic equation (\ref{frw-10}) will admit one real root and two complex roots; otherwise the cubic equation (\ref{frw-10}) will admit all real roots. 

As a result, the corresponding graviton terms can be defined to be
\begin{equation}
{\cal L}_M = 2 \hat\Sigma^2 \left(\alpha_4 \hat\Sigma^2 +4\alpha_3 \hat\Sigma +6 \right),
\end{equation}
which implies the following effective cosmological constant $\Lambda_M$:
\begin{equation}
\Lambda_M \equiv - m_g^2\frac{{\cal L}_M}{2} =-m_g^2 \hat\Sigma^2 \left(\alpha_4 \hat\Sigma^2 +4\alpha_3 \hat\Sigma +6 \right).
\end{equation}

Now, we would like to discuss the special case:
\begin{equation}
\alpha_4 \hat\Sigma^2+3\alpha_3 \hat\Sigma+3 =0,
\end{equation}
which will be solved to give
\begin{equation}
\hat \Sigma = \frac{-3\alpha_3 \pm \sqrt{9\alpha_3^2 -12\alpha_4}}{2\alpha_4}
\end{equation}
along with a condition that $\alpha_3^2 > 4\alpha_4/3$. It is appears that  $\alpha_5$ will no longer be free due to Eq. (\ref{frw-11}). Indeed, the corresponding $\alpha_5$ will be determined in terms of the other parameters as follows
\begin{eqnarray} \label{special-a5}
\alpha_5& =&- \frac{3\alpha_4 \hat \Sigma^2 +3 \alpha_3 \hat \Sigma +1}{\hat \Sigma^3} \nonumber\\
&=&\frac{8\alpha _4^2 \left[\left(9 \alpha _3^2-8 \alpha _4\right)\mp3 \alpha _3 \sqrt{9 \alpha _3^2-12 \alpha _4}  \right] }{\left(3 \alpha _3\mp\sqrt{9 \alpha _3^2-12 \alpha _4}\right)^3}. \nonumber\\
\end{eqnarray}
 In addition, given the above solutions $\hat\gamma$ turns out to be positively arbitrary. Consequently,  the corresponding ${\cal L}_M$ turns out to be
\begin{eqnarray}
{\cal L}_M &=&2 \hat\Sigma^2 \left(\alpha_3 \hat\Sigma +3 \right)\nonumber\\
 &=&  -\frac{3}{\alpha_4^3} \biggl[9\alpha_3^4 -18\alpha_3^2 \alpha_4 +6\alpha_4^2  \nonumber\\
&&  \mp \alpha_3 \left(3\alpha_3 ^2-4\alpha_4 \right) \sqrt{9\alpha_3^2 -12\alpha_4}\biggr].
\end{eqnarray}
Hence, the corresponding effective cosmological constant can be defined to be
\begin{eqnarray} \label{frw-special}
\Lambda_M &=& \frac{3m_g^2}{2\alpha_4^3} \biggl[9\alpha_3^4 -18\alpha_3^2 \alpha_4 +6\alpha_4^2  \nonumber\\
&&  \mp \alpha_3 \left(3\alpha_3 ^2-4\alpha_4 \right) \sqrt{9\alpha_3^2 -12\alpha_4}\biggr],
\end{eqnarray}
 which is independent of $\alpha_5$. Moreover, the effective cosmological constant $\Lambda_M$, can be either negative or positive definite depending on the value of $\alpha_3$ and $\alpha_4$. It is worth noting that the value of effective cosmological constant in the special case as shown in Eq. (\ref{frw-special}) is identical to $\Lambda^0_M$ defined in Eq. (\ref{lambda1}), which has been investigated in four-dimensional dRGT theory ~\cite{WFK}. In other words, by fine-tuning $\alpha_5$ such that it satisfies the condition (\ref{special-a5}) then the effective cosmological constant from the graviton terms in five-dimensional FLRW spacetime will coincide with that in four-dimensional FLRW spacetime, where ${\cal L}_5$ disappears automatically as claimed before.

Given the result that $m_g^2{\cal L}_M =-2 {\Lambda}_M$, the modified Einstein equations (\ref{Einstein-5d-reduced}) in the FLRW spacetime now become
\begin{equation}
\left({R_{\mu\nu}-\frac{1}{2}Rg_{\mu\nu}}\right) +\Lambda_M g_{\mu\nu}=0.
\end{equation}
If we set the scale factor of physical metric as 
\begin{equation}
a_1(t) = \exp\left[\tilde\alpha t\right]
\end{equation} 
along with $N_1(t)=1$, then we can solve the Einstein equations with the positive $\Lambda_M$ to obtain the following solution:
\begin{equation}
\tilde\alpha =\sqrt{\frac{\Lambda_M}{6}},
\end{equation}
which seems to be the de Sitter solution in five dimensions. Note again that the four-dimensional dRGT theory also admits the de Sitter metric as its cosmological solution if both physical and fiducial metrics are assumed to be FLRW one as mentioned in Sec. ~\ref{sec2}. 
\section{Five-dimensional Bianchi type I metrics} \label{sec5}
In five-dimensional spacetime, the Bianchi type I physical and fiducial metrics can be taken to be  ~\cite{bianchi-I,WFK,5d-bianchi}
\begin{eqnarray}
\label{eq7}
 ds_{5d}^2 (g_{\mu\nu}) &= & -N_1^2(t)dt^2+\exp\left[{2\alpha_1(t)-4\sigma_1(t)}\right]dx^2\nonumber \\  &&+\exp\left[{2\alpha_1(t)+2\sigma_1(t)}\right]\left({dy^2+dz^2}\right) \nonumber \\ 
&& +\exp\left[{2\beta_1(t)}\right] du^2 , 
\end{eqnarray}
\begin{eqnarray}
 \label{eq8}
 ds_{5d}^2 (Z_{\mu\nu})  &= &  -N_2^2(t)dt^2+\exp\left[{2\alpha_2(t)-4\sigma_2(t)}\right]dx^2\nonumber \\  &&+\exp\left[{2\alpha_2(t)+2\sigma_2(t)}\right]\left({dy^2+dz^2}\right) \nonumber \\
&& +\exp\left[{2\beta_2(t)}\right] du^2 , 
\end{eqnarray}
where $\beta_i$'s ($i=1-2$) are additional scale factors associated with the fifth dimension $u$ ~\cite{5d-bianchi}. In addition, the  St\"uckelberg scalar fields have been taken to be in the unitary gauge, $\phi^a =x^a$, i.e., $\partial_\mu \phi^a =\delta_\mu^a$ ~\cite{KNT,RGT2,TMN,RGT1,bianchi-I,WFK}. 

In order to compare with the results in four-dimensional case, where $\beta_i$'s associated with the fifth dimension have not been introduced, we will use the notations as used in Ref.~ \cite{WFK}. In particular, we will define the following variables in five-dimensional spacetime:
\begin{eqnarray}
\left[{\cal K} \right]^n &=& \left(5-\gamma-A-2B-C\right)^n, \nonumber\\
\left[{\cal K}^n \right] &=& \left(1-\gamma \right)^n +\left(1-A\right)^n +2\left(1-B\right)^n +\left(1-C\right)^n, \nonumber\\
\label{def-of-gamma}
\gamma &=& \frac{N_2}{N_1};~A=\epsilon \eta^{-2}; ~ B=\epsilon \eta; ~C= \exp\left[\beta_2-\beta_1\right], \nonumber\\
\epsilon &= &\exp\left[\alpha_2 -\alpha_1 \right]; ~ \eta =\exp\left[\sigma_2-\sigma_1\right].
\end{eqnarray}
Given these definitions, we are able to define the following graviton terms ${\cal L}_i$ such as
\begin{eqnarray}
{\cal L}_2 &=&2 \Bigl[B \left(2A+B\right) +\left(\gamma-3\right)\left(A+2B\right) +3\left(2-\gamma\right) \Bigr] \nonumber\\
&&+ 2\left(A+2B+\gamma-4\right)\left(C-1\right),
\end{eqnarray}
\begin{eqnarray}
{\cal L}_3 &=& -2 \Bigl[ AB^2   + \left(\gamma-2 \right)B \left(2A+B \right) \nonumber\\
&& + \left(3-2\gamma \right) \left(A+2B\right) +3\gamma-4 \Bigr] -2  \Bigl[ B \left(2A+B \right)  \nonumber\\
&& +\left(\gamma-3\right)\left(A+2B\right) +3 \left(2-\gamma \right) \Bigr]\left(C-1\right), 
\end{eqnarray}
\begin{eqnarray}
{\cal L}_4&=& 2 \left(\gamma-1\right) \left(A-1 \right) \left(B-1\right)^2 \nonumber\\
&& +2  \Bigl[ AB^2  + \left(\gamma-2 \right)B \left(2A+B \right)  \nonumber\\
&&+ \left(3-2\gamma \right) \left(A+2B\right) +3\gamma-4 \Bigr]  \left(C-1\right), 
\end{eqnarray}
\begin{eqnarray}
{\cal L}_5&=&-2 \left(\gamma-1 \right) \left(A-1\right) \left(B-1\right)^2 \left(C-1\right).
\end{eqnarray}
Hence, taking a summation of all graviton terms leads to
\begin{eqnarray} \label{explicit-LM}
{\cal L}_M  = {\cal L}_M^0+{\cal L}_M^C, 
\end{eqnarray}
where
\begin{eqnarray} \label{explicit-LM0}
{\cal L}_M^0 &=& 2\Bigl[{\left({\gamma\alpha_4-\gamma_3}\right) AB^2 +\left({\gamma_2-\gamma\gamma_3}\right)B\left({2A+B}\right) }\nonumber\\ 
&& {+\left({\gamma\gamma_2-\gamma_1}\right)\left({A+2B}\right) -\gamma \gamma_1 +\left({3\gamma_1-3\gamma_2+\gamma_3}\right) }\Bigr]  , \nonumber\\ \\
 \label{explicit-LMC}
{\cal L}_M^C &=& 2 \Bigl\{ \alpha_4 AB^2 -\alpha_5 \left(\gamma-1\right) \left(A-1\right) \left(B-1\right)^2  \nonumber\\
&&  - \left[\gamma_3- \left(\gamma-1\right)\alpha_4 \right] B \left(2A+B\right)  \nonumber\\
&&  +\left[\gamma_2 -\left(\gamma -1\right)  \left(\gamma_3+\alpha_4 \right) \right] \left(A+2B\right)   \nonumber\\
&& + \left(\gamma-1\right) \left(3\gamma_3 +1 \right) -\gamma_1 \Bigr\} \left(C-1\right).
\end{eqnarray}
It is straightforward to check that if we take the limit, $\beta_i \to 0$ ($C\to 1$), then ${\cal L}_5 \to 0$, ${\cal L}_M^C \to 0$, and ${\cal L}_{2,3,4,M}$ all reduce to that defined in four-dimensional spacetime framework ~\cite{WFK}. Moreover,  the ${\cal L}_i$ ($i=2-5$) and ${\cal L}_M$ all reduce to that defined in the previous section in the isotropic FLRW limit, $\sigma_1=\sigma_2=0$, $\beta_i =\alpha_i$, and $A=B=C$.
\subsection{Constraint equations}
Following the method in Ref. ~\cite{WFK}, we solve the constraint equations associated with the scale factors of the fiducial metric, $N_2$, $\alpha_2$, $\sigma_2$, and $\beta_2$ in order to show constant-like property of graviton terms. In particular, these constraint equations are the Euler-Lagrange equations:
\begin{equation} \label{Euler}
\frac{\partial {\cal L}_M}{\partial N_2}=0; ~ \frac{\partial {\cal L}_M}{\partial \alpha_2}=0; ~\frac{\partial {\cal L}_M}{\partial \sigma_2}=0; ~\frac{\partial {\cal L}_M}{\partial \beta_2}=0.
\end{equation}
Here, these constraint equations are derived following the fact that the graviton Lagrangian ${\cal L}_M$ does not involve any time derivative of $N_2$, $\alpha_2$, $\sigma_2$, and $\beta_2$ as shown in Eq. (\ref{explicit-LM}). As a result, it is straightforward to show that the Euler-Lagrange equations (\ref{Euler}) can be reduced to
\begin{equation} \label{Euler-1}
\frac{\partial {\cal L}_M}{\partial \gamma}=0; ~ \frac{\partial {\cal L}_M}{\partial A}=0; ~\frac{\partial {\cal L}_M}{\partial B}=0; ~\frac{\partial {\cal L}_M}{\partial C}=0,
\end{equation}
where $\gamma$, $A$, $B$, and $C$ are functions of $N_2$, $\alpha_2$, $\sigma_2$, and $\beta_2$, respectively, as  defined in Eq. (\ref{def-of-gamma}). Given the graviton Lagrangian ${\cal L}_M$ in Eq. (\ref{explicit-LM}), we are able to obtain the following algebraic constraint equations:
\begin{eqnarray} \label{bianchi-constraint-1}
&&\alpha_4 AB^2-\gamma_3B\left({2A+B}\right) +\gamma_2\left({A+2B}\right)-\gamma_1 \nonumber\\
&&+\Bigl[  \alpha_4 B\left(2A+B\right) -\alpha_5 \left(A-1\right)\left(B-1\right)^2 \nonumber\\
&&  -\left(\gamma_3+\alpha_4\right) \left(A+2B\right) +3\gamma_3 +1 \Bigr] \left(C-1\right)=0, \nonumber\\
\end{eqnarray}
\begin{eqnarray}\label{bianchi-constraint-2}
&&\left({\gamma \alpha_4  -\gamma_3}\right)B^2+2\left({\gamma_2-\gamma \gamma_3}\right)B-\gamma_1+\gamma\gamma_2\nonumber\\
 &&+ \Bigl\{ \alpha_4 B^2  -\alpha_5 \left(\gamma-1\right) \left(B-1\right)^2- 2\left[\gamma_3 - \left(\gamma-1\right)\alpha_4\right]B  \nonumber\\
&&  - \left(\gamma-1\right) \left(\gamma_3+\alpha_4\right) +\gamma_2 \Bigr\} \left(C-1\right)=0, 
\end{eqnarray}
\begin{eqnarray}\label{bianchi-constraint-3}
&&\left({\gamma \alpha_4  -\gamma_3}\right)AB +\left({\gamma_2-\gamma \gamma_3}\right)\left({A+B}\right)-\gamma_1+\gamma \gamma_2 \nonumber\\
 && +\Bigl\{ \alpha_4 AB-\alpha_5 \left(\gamma-1\right) \left(A-1\right)\left(B-1\right) \nonumber\\
&& - \left[\gamma_3 - \left(\gamma-1\right)\alpha_4\right] \left(A+B\right)  \nonumber\\
&&   - \left(\gamma-1\right) \left(\gamma_3+\alpha_4\right) +\gamma_2 \Bigr\} \left(C-1\right)=0, 
\end{eqnarray}
\begin{eqnarray}\label{bianchi-constraint-4}
&&\alpha_4 AB^2-\gamma_3B\left({2A+B}\right) +\gamma_2\left({A+2B}\right)-\gamma_1 \nonumber\\
&&+\Bigl[  \alpha_4 B\left(2A+B\right) -\alpha_5 \left(A-1\right)\left(B-1\right)^2 \nonumber\\
&&  -\left(\gamma_3+\alpha_4\right) \left(A+2B\right) +3\gamma_3 +1 \Bigr] \left(\gamma-1\right)=0. \nonumber\\
\end{eqnarray}

It is straightforward to check that under the limit, $C\to 1$, all above algebraic constraint equations reduce to that found in the four-dimensional dRGT theory ~\cite{WFK}. 

As a result, solving Eqs. (\ref{bianchi-constraint-2}) and (\ref{bianchi-constraint-3}) leads to two possible solutions:
\begin{eqnarray}\label{bianchi-constraint-5}
B&= & A, \\
\label{bianchi-constraint-6}
C&=&1-\frac{\left(\gamma \alpha_4 -\gamma_3\right)B +\gamma_2 -\gamma \gamma_3}{\alpha_4 B - \alpha_5 \left(\gamma-1\right) \left(B-1\right) +\alpha_4\left(\gamma-1\right)-\gamma_3  }. \nonumber\\
\end{eqnarray}
On the other hand, we obtain from Eqs. (\ref{bianchi-constraint-1}) and (\ref{bianchi-constraint-4})  two possible solutions: 
\begin{eqnarray}\label{bianchi-constraint-7}
&& \alpha_4 AB^2 -\gamma_3B\left({2A+B}\right)+\gamma_2\left({A+2B}\right)-\gamma_1 =0, \nonumber\\ \\
\label{bianchi-constraint-8}
&&\gamma =C.
\end{eqnarray}
It is apparent that we now have four possible cases of the solutions listed in two sets:  Eqs. (\ref{bianchi-constraint-5})-(\ref{bianchi-constraint-6}) and Eqs. (\ref{bianchi-constraint-7})-(\ref{bianchi-constraint-8}). Next, we will study whether these solutions lead to physical solutions of the constraint equations. Additionally, we will compute the corresponding values of massive graviton terms ${\cal L}_M$ once the final solutions of the constraint equations are figured out. 
\subsubsection{Case 1}
First, we consider a case, in which the solutions shown in Eqs. (\ref{bianchi-constraint-5}) and (\ref{bianchi-constraint-7}):
\begin{eqnarray}
B &=&  A, \nonumber\\
\alpha_4 AB^2 -\gamma_3B\left({2A+B}\right)+\gamma_2\left({A+2B}\right)-\gamma_1 &=&0, \nonumber
\end{eqnarray} 
are chosen to solve the constraint equations. In particular, for the solution $B=A$ the equation (\ref{bianchi-constraint-7}) can be factorized as
\begin{eqnarray}
\left(A-1\right) \left[\alpha_4 A^2 - \left(3\alpha_3+2\alpha_4 \right) A +3\alpha_3 +\alpha_4+3 \right]=0, \nonumber\\
\end{eqnarray}
where $\gamma_{1,2,3}$'s defined in Eq. (\ref{def-gamma}) have been used in order to derive the above equation. As a result, this equation admits non-trivial solutions:
\begin{equation}\label{bianchi-case1-sol1}
A=1+\frac{3\alpha_3 \pm \sqrt{3 \left(3\alpha_3^2 -4\alpha_4 \right)}}{2\alpha_4},
\end{equation}
here we have ignored a trivial solution, $A=1$.  As a result, plugging the solution (\ref{bianchi-case1-sol1}) into Eq. (\ref{bianchi-constraint-1}) or (\ref{bianchi-constraint-4}) leads to a relation between $\alpha_5$ and the other $\alpha_3$ and $\alpha_4$ as follows
\begin{eqnarray}\label{bianchi-case1-sol2}
\alpha_5 &=&- \frac{3\alpha_4 \left(1-A\right)^2+3\alpha_3 \left(1-A\right) +1}{\left(1-A\right)^3} \nonumber\\
&=& \frac{8\alpha _4^2 \left[\left(9 \alpha _3^2-8 \alpha _4\right)\pm3 \alpha _3 \sqrt{9 \alpha _3^2-12 \alpha _4}  \right] }{\left(3 \alpha _3\pm\sqrt{9 \alpha _3^2-12 \alpha _4}\right)^3}, \nonumber\\
\end{eqnarray}
which is identical to that defined in Eq. (\ref{special-a5}). Here we have also neglected  trivial solutions, $C=1$ and $\gamma=1$.  Furthermore, given the solutions (\ref{bianchi-case1-sol1}) and (\ref{bianchi-case1-sol2}) we are able to reduce both Eqs. (\ref{bianchi-constraint-2}) and (\ref{bianchi-constraint-3}) to
\begin{equation}\label{bianchi-case1-sol3}
\left(C-A\right) \left(\gamma -A \right)=0.
\end{equation}
It turns out that solutions to Eq. (\ref{bianchi-case1-sol3}) are given by
\begin{equation}\label{bianchi-case1-sol4}
C=A ~\text{~for arbitrary~}~ \gamma ,
\end{equation}
or
\begin{equation}\label{bianchi-case1-sol5}
\gamma =A ~\text{~for arbitrary~}~ C .  
\end{equation} 
Interestingly, for either the solution listed in Eq. (\ref{bianchi-case1-sol4}) or that in Eq. (\ref{bianchi-case1-sol5}), ${\cal L}_M^C$ shown in Eq. (\ref{explicit-LMC}) always vanishes. Hence, we arrive at a result:
\begin{eqnarray}
{\cal L}_M& =& {\cal L}_M^0 = -\frac{3}{\alpha_4^3} \biggl[9\alpha_3^4 -18\alpha_3^2\alpha_4 +6\alpha_4^2 \nonumber\\
&&  \pm \alpha_3 \left(3\alpha_3^2 -4\alpha_4\right)\sqrt{9 \alpha _3^2-12 \alpha _4}\biggr].
\end{eqnarray}
The values of an effective cosmological constant associated with the massive graviton terms can therefore be determined to be
\begin{eqnarray}
\Lambda_M &=& \frac{3m_g^2}{2\alpha_4^3}\biggl[9\alpha_3^4 -18\alpha_3^2\alpha_4 +6\alpha_4^2  \nonumber\\
&& \pm \alpha_3 \left(3\alpha_3^2 -4\alpha_4\right)\sqrt{9 \alpha _3^2-12 \alpha _4}\biggr],
\end{eqnarray}
which are identical to that found in the four-dimensional dRGT theory ~\cite{WFK}. To end this case, we would like to remark that the field parameter $\alpha_5$ associated with the existence of ${\cal L}_5$ cannot be arbitrary but be constrained by the other $\alpha_3$ and $\alpha_4$ as shown in Eq. (\ref{bianchi-case1-sol2}). In addition, $\alpha_3^2>4\alpha_4/3$ is also required to make the effective cosmological constant $\Lambda_M$ real definite.  
\subsubsection{Case 2}
Second, we consider another case, in which the solutions shown in Eqs. (\ref{bianchi-constraint-5}) and (\ref{bianchi-constraint-8}):
\begin{eqnarray}
B=  A, ~\gamma = C, \nonumber
\end{eqnarray} 
are selected to solve the constraint equations. As a result, we get two following equations from the constraint equations (\ref{bianchi-constraint-1}) and (\ref{bianchi-constraint-2}) such as
\begin{eqnarray}\label{bianchi-case2-sol1}
&&\left[\alpha_4 \left(A-1\right)^2-3\alpha_3 \left(A-1\right)+3\right] \left(A-1\right) - \left[\alpha_5 \left(A-1\right)^3 \right. \nonumber\\
&& \left. - 3\alpha_4 \left(A-1\right)^2+3\alpha_3 \left(A-1\right)-1\right] \left(C-1\right)=0,
\end{eqnarray}
\begin{eqnarray}\label{bianchi-case2-sol2}
&&\alpha_4 \left(A-1\right)^2 \left(C-1\right)-\alpha_3 \left(A-1\right)\left[A-1 +2 \left(C-1\right)\right] \nonumber\\
&& +2\left(A-1\right)+\left(C-1\right) - \Bigl\{ \alpha_5 \left(A-1\right)^2\left(C-1\right) \nonumber\\
&&  -\alpha_4 \left(A-1\right) \left[A-1+2\left(C-1\right)\right] \nonumber\\
&& +\alpha_3 \left[2\left(A-1\right)+C-1\right]-1 \Bigr\}  \left(C-1\right)=0,
\end{eqnarray}
which can be combined to give
\begin{eqnarray}\label{bianchi-case2-sol3}
&&\Bigl\{\alpha_4 \left(A-1\right)^2 -2\alpha_3 \left(A-1\right)+1 -\bigl[\alpha_5 \left(A-1\right)^2  \nonumber\\
&&  -2\alpha_4 \left(A-1\right)+\alpha_3 \bigr]\left(C-1\right) \Bigr\} \left(C-A\right) =0. 
\end{eqnarray}
Hence, two possible solutions of Eq. (\ref{bianchi-case2-sol3}) read
\begin{eqnarray}\label{bianchi-case2-sol4}
C&=&A,\\
\label{bianchi-case2-sol5}
C&=&1+\frac{\alpha_4 \left(A-1\right)^2 -2\alpha_3 \left(A-1\right)+1}{\alpha_5 \left(A-1\right)^2-2\alpha_4 \left(A-1\right)+\alpha_3}.
\end{eqnarray}
As a result, for the first solution of Eq. (\ref{bianchi-case2-sol3}) as shown in Eq. (\ref{bianchi-case2-sol4}) the constraint equation (\ref{bianchi-constraint-1}) or (\ref{bianchi-constraint-2})  reduces to a cubic equation of a variable ${\hat A} \equiv 1-A$:
\begin{equation}\label{bianchi-case2-sol6}
 \alpha_5 {\hat A}^3 +4\alpha_4 {\hat A}^2  +6\alpha_3 {\hat A} +4 =0.
\end{equation}
Additionally, the corresponding graviton terms become
\begin{equation}\label{bianchi-case2-sol7}
{\cal L}_M = 2 {\hat A}^2 \left[\alpha_4 {\hat A}^2 +4\alpha_3 {\hat A}+6\right].
\end{equation}
As a result, the equation (\ref{bianchi-case2-sol6}) can be solved to give three possible roots of $\hat A$, which can be either real or complex definite. However, it is noted that ${\cal L}_M$ should be real (positive or negative) definite. Consequently, the variable ${\hat A}$ should be real definite, too.

For the other solution shown in Eq. (\ref{bianchi-case2-sol5}), the corresponding equation of $\hat A$ turns out to be a quartic equation:
\begin{eqnarray}\label{bianchi-case2-sol8}
&& \left(\alpha_4^2 -\alpha_3 \alpha_5 \right){\hat A}^4 +2 \left(\alpha_3 \alpha_4 -\alpha_5\right){\hat A}^3 \nonumber\\
&& +\left(3\alpha_3^2 -2\alpha_4 \right){\hat A}^2 +2\alpha_3 \hat A+1=0
\end{eqnarray}
along with the following graviton terms defined to be
\begin{eqnarray}
{\cal L}_M &= &-2  \left[ \left(\alpha_4^3 -\alpha_5^2\right){\hat A}^6 +\left( 3\alpha_3^2 \alpha_5+3\alpha_3 \alpha_4^2  -4\alpha_4 \alpha_5 \right){\hat A}^5 \right. \nonumber\\
&& \left.  +3 \left(2\alpha_3^2 \alpha_4 -\alpha_4^2+ 2\alpha_3 \alpha_5\right){\hat A}^4 \right. \nonumber\\
&& \left. -\left(\alpha_3^3 -10\alpha_3 \alpha_4 -4\alpha_5 \right){\hat A}^3 - \left(\alpha_3^2 -7\alpha_4\right){\hat A}^2 \right. \nonumber\\
&& \left.  -\alpha_3 {\hat A}-1\right] \times \left[\alpha_5 {\hat A}^2 +2\alpha_4 {\hat A} +\alpha_3 \right]^{-2}.
\end{eqnarray}
As a result, four possible values of $\hat A$, which are either real or complex definite, can be figured out from the following quartic Eq. (\ref{bianchi-case2-sol8}). Note again that $\hat A$ must be real in order to have the real effective cosmological constant $\Lambda_M = -m_g^2{\cal L}_M/2$.  
\subsubsection{Case 3}
Next, we consider the case, in which the solutions shown in Eqs. (\ref{bianchi-constraint-6}) and (\ref{bianchi-constraint-7}):
\begin{eqnarray}
&&C=1-\frac{\left(\gamma \alpha_4 -\gamma_3\right)B +\gamma_2 -\gamma \gamma_3}{\alpha_4 B - \alpha_5 \left(\gamma-1\right) \left(B-1\right)+\alpha_4\left(\gamma-1\right)-\gamma_3  } , \nonumber\\
&& \alpha_4 AB^2 -\gamma_3B\left({2A+B}\right)+\gamma_2\left({A+2B}\right)-\gamma_1 =0, \nonumber
\end{eqnarray} 
are considered to solve the constraint equations. From the second equation, we can define $A$ as a function of $B$ as follows
\begin{equation}\label{bianchi-case3-sol1}
A=\frac{\gamma_3 B^2 -2\gamma_2 B +\gamma_1}{\alpha_4 B^2 -2\gamma_3 B +\gamma_2}.
\end{equation}
As a result,  both Eqs. (\ref{bianchi-constraint-1}) and (\ref{bianchi-constraint-4}) reduce, due to the solution (\ref{bianchi-constraint-7}), to
\begin{eqnarray}\label{bianchi-case3-sol2}
&& \left(\gamma_3+\alpha_4\right) \left(A+2B\right) -\alpha_4 B\left(2A+B\right) \nonumber\\
&& +\alpha_5 \left(A-1\right)\left(B-1\right)^2 -3\gamma_3 -1=0,
\end{eqnarray}
here the trivial solutions, $C=1$ and $\gamma=1$, have been neglected. Now, inserting the relation between $A$ and $B$ as shown in Eq. (\ref{bianchi-case3-sol1}) into Eq. (\ref{bianchi-case3-sol2}) gives a quartic equation of  ${\hat B} \equiv 1-B$:
\begin{eqnarray}\label{bianchi-case3-sol3}
&&\left(\alpha_4^2 -\alpha_3 \alpha_5\right) {\hat B}^4 +2\left(\alpha_3 \alpha_4-\alpha_5\right){\hat B}^3 \nonumber\\
&&+\left(3\alpha_3^2 -2\alpha_4 \right){\hat B}^2 + 2\alpha_3 {\hat B}+1 =0.
\end{eqnarray}
It is straightforward to show that this equation (of $\hat B$) is exactly the equation (\ref{bianchi-case2-sol8}) of ${\hat A}$ in the previous case, where $A=B$ and $\gamma=C$. It is apparent that coefficients of Eq. (\ref{bianchi-case3-sol3}) depend only on the field parameters, $\alpha_3$, $\alpha_4$, and $\alpha_5$. Hence, solutions of this equation can be determined up to three parameters,  $\alpha_3$, $\alpha_4$, and $\alpha_5$. Once values of $\hat B$ (or $B$) are calculated, the corresponding values of $A$ will be worked out, according to Eq. (\ref{bianchi-case3-sol1}). 

On the other hand, plugging the solution (\ref{bianchi-constraint-6}) into either Eq. (\ref{bianchi-constraint-2}) or Eq. (\ref{bianchi-constraint-3}) leads to a quadratic equation of ${\hat \gamma} \equiv 1-\gamma$:
\begin{eqnarray}\label{bianchi-case3-sol4}
&& \left[\left(\alpha_4^2 -\alpha_3 \alpha_5\right) {\hat B}^2 + \left(\alpha_3 \alpha_4-\alpha_5\right){\hat B}+\alpha_3^2-\alpha_4 \right]{\hat \gamma}^2 \nonumber\\
&& -\left[\left(\alpha_5-\alpha_3 \alpha_4 \right) {\hat B}^2 -\alpha_3^2 {\hat B} -\alpha_3 \right]{\hat \gamma} \nonumber\\
&& +\left(\alpha_3^2 -\alpha_4\right){\hat B}^2 +\alpha_3 {\hat B}+1 =0.
\end{eqnarray}
As a result, once Eq. (\ref{bianchi-case3-sol3}) is solved, we will have values of ${\hat B}$ defined in terms of the field parameters, $\alpha_3$, $\alpha_4$, and $\alpha_5$. Then, all coefficients of Eq. (\ref{bianchi-case3-sol4}) will be evaluated such that values of $\hat\gamma$ will be figured out consistently. Additionally, values of $C$ will also be defined, according to the relation as shown in Eq. (\ref{bianchi-constraint-6}). Of course, once all variables $A$, $B$, $C$, and $\gamma$ are solved, the massive graviton term ${\cal L}_M$ shown in Eq. (\ref{explicit-LM}) will be determined directly.
\subsubsection{Case 4}
Finally, we consider the fourth case, in which the solutions shown in Eqs. (\ref{bianchi-constraint-6}) and (\ref{bianchi-constraint-8}):
\begin{eqnarray}
C &=& 1-\frac{\left(\gamma \alpha_4 -\gamma_3\right)B +\gamma_2 -\gamma \gamma_3}{\alpha_4 B - \alpha_5 \left(\gamma-1\right) \left(B-1\right)+\alpha_4\left(\gamma-1\right) -\gamma_3  }, \nonumber\\
C& =& \gamma ,\nonumber
\end{eqnarray}
 are used to solve the constraint equations. It is straightforward to obtain, from these two solutions, a quadratic equation of $\hat \gamma$:
\begin{equation}\label{bianchi-case4-sol1}
\left(\alpha_5 {\hat B}+\alpha_4 \right) {\hat \gamma}^2 -2\left(\alpha_4 {\hat B} +\alpha_3 \right) {\hat \gamma} +\alpha_3 {\hat B}+1=0,
\end{equation}
where ${\hat B} =1-B$ and ${\hat \gamma}=1-\gamma$. Similar to the third case, either Eq. (\ref{bianchi-constraint-2}) or Eq. (\ref{bianchi-constraint-3}) can be reduced to another quadratic equation of ${\hat \gamma}$:
\begin{eqnarray}\label{bianchi-case4-sol2}
&& \left[\left(\alpha_4^2 -\alpha_3 \alpha_5\right) {\hat B}^2 + \left(\alpha_3 \alpha_4-\alpha_5\right){\hat B}+\alpha_3^2-\alpha_4 \right]{\hat \gamma}^2 \nonumber\\
&&-\left[\left(\alpha_5-\alpha_3 \alpha_4 \right) {\hat B}^2 -\alpha_3^2 {\hat B} -\alpha_3 \right]{\hat \gamma} \nonumber\\
&& +\left(\alpha_3^2 -\alpha_4\right){\hat B}^2 +\alpha_3 {\hat B}+1 =0,
\end{eqnarray}
with the help of the solution (\ref{bianchi-constraint-6}). Equating Eqs. (\ref{bianchi-case4-sol1}) and (\ref{bianchi-case4-sol2}) leads to a set of solutions: $\alpha_3 =3$,  $\alpha_4 =9$, $\alpha_5 = 27$, and ${\hat B}=-{1}/{3}$ (or equivalently $B=4/3$). However, these solutions do not satisfy both constraint equations (\ref{bianchi-constraint-1}) and (\ref{bianchi-constraint-4}). Hence, we conclude that there is no any solution to the constraint equations in this case. 
\subsection{Einstein field equations}
Armed with the effective cosmological constant $\Lambda_M=-m_g^2{\cal L}_M/2$, whose values have been defined  in the above subsection, we arrive at the five-dimensional  Einstein field equations as
\begin{equation} \label{bianchi-I-Einstein-1}
\left({R_{\mu\nu}-\frac{1}{2}Rg_{\mu\nu}}\right) +\Lambda_M g_{\mu\nu}=0.
\end{equation}
As a result, given the physical Bianchi type I metric (\ref{eq7}) with $N_1(t)=1$ the following component equations of Eq. (\ref{bianchi-I-Einstein-1}) turn out to be
\begin{eqnarray}
 \label{bianchi-I-Einstein-2}
3\left(\dot\alpha_1^2 -\dot\sigma_1^2+\dot\alpha_1 \dot\beta_1  \right) &=& \Lambda_M,\\
 \label{bianchi-I-Einstein-3}
3\left(\ddot\alpha_1+2\dot\alpha_1^2 +\dot\sigma_1^2\right) &=&\Lambda_M,\\
 \label{bianchi-I-Einstein-4}
\ddot\sigma_1 + \dot\sigma_1 \left(3\dot\alpha_1 + \dot\beta_1 \right)& =&0,\\
 \label{bianchi-I-Einstein-5}
\ddot\beta_1 -2\dot\alpha_1^2+2\dot\sigma_1^2+\dot\beta_1^2+\dot\alpha_1 \dot\beta_1 &=&0.
\end{eqnarray}
It is apparent that we have ended up with a set of four differential equations of three independent variables. In addition, Eq. (\ref{bianchi-I-Einstein-2}) is the Friedmann equation acting as the constraint equation. Now, we will try to find more convenient relations between these variables from the above equations. First, we obtain from Eqs. (\ref{bianchi-I-Einstein-2}) and (\ref{bianchi-I-Einstein-5}) that
\begin{equation} \label{bianchi-I-Einstein-6}
3\ddot\beta_1 +3\dot\beta_1^2 +9\dot\alpha_1 \dot\beta_1 =2\Lambda_M.
\end{equation}
In addition, combining both Eqs. (\ref{bianchi-I-Einstein-3}) and (\ref{bianchi-I-Einstein-5}) leads to
\begin{equation}\label{bianchi-I-Einstein-7}
6\ddot\alpha_1 -3\ddot\beta_1 +18\dot\alpha_1^2-3\dot\alpha_1 \dot\beta_1 -3\dot\beta_1^2 =2\Lambda_M.
\end{equation}
As a result, Eq.  (\ref{bianchi-I-Einstein-7}) can be reduced to
\begin{equation}\label{bianchi-I-Einstein-8}
18\left(\ddot\alpha_1 +3\dot\alpha_1^2\right)-6\left(\ddot\beta_1 +\dot\beta_1^2\right)=8\Lambda_M,
\end{equation}
with the help of Eq. (\ref{bianchi-I-Einstein-6}). By introducing additional variables such as~\cite{WFK} $V_1=\exp[3\alpha_1]$ and $V_2=\exp[\beta_1]$, we will be able to rewrite Eq. (\ref{bianchi-I-Einstein-8}) as follows
\begin{equation}\label{bianchi-I-Einstein-9}
\frac{\ddot V_1}{V_1} -\frac{\ddot V_2}{V_2} =\frac{4\Lambda_M}{3}.
\end{equation}
Furthermore, if we assume
\begin{equation}
\frac{\ddot V_2}{V_2} = V_0 \frac{\ddot V_1}{V_1} , 
\end{equation}
where $V_0$ is a constant, then the above equation (\ref{bianchi-I-Einstein-9}) will be reduced to a linear differential equation in $V_1$:
\begin{equation}\label{bianchi-I-Einstein-10}
\ddot V_1 = 9 \tilde H_1^2 V_1 ,
\end{equation}
where $\tilde H_1^2 =4 H_1^2/9(1-V_0)$ along with $H_1^2 \equiv \Lambda_M/3$ as a Hubble constant and a requirement $0<V_0<1$. As a result, an explicit solution of Eq. (\ref{bianchi-I-Einstein-10}) is given by  ~\cite{WFK}
\begin{eqnarray}\label{bianchi-I-Einstein-12}
V_1 &\equiv& \exp \left[3\alpha_1\right] \nonumber\\
&=&\exp\left[3\alpha_0\right] \left[\cosh \left(3\tilde H_1 t\right)+\frac{\dot\alpha_0}{\tilde H_1} \sinh \left(3\tilde H_1 t \right) \right], \nonumber\\
\end{eqnarray}
where $\alpha_0 \equiv \alpha_1(t=0)$ and $\dot\alpha_0 \equiv \dot\alpha_1(t=0)$ are initial values. Similarly, for $V_2$ we obtain the following linear differential equation:
\begin{equation}\label{bianchi-I-Einstein-13}
\ddot V_2 = 9 {\bar H}_1^2 V_2,
\end{equation}
where $\bar H_1^2 = V_0 \tilde H_1^2$. Therefore, this equation can be solved to give a solution:
\begin{eqnarray}\label{bianchi-I-Einstein-14}
V_2 &\equiv& \exp \left[\beta_1\right] \nonumber\\
&=&\exp\left[\beta_0\right] \left[\cosh \left(3\bar H_1 t\right)+\frac{\dot\beta_0}{3\bar H_1} \sinh \left(3\bar H_1 t \right) \right], \nonumber\\
\end{eqnarray}
where $\beta_0 = \beta_1(t=0)$ and $\dot\beta_0 =\dot\beta_1(t=0)$ acting as initial values. For the last variable, $\sigma_1$, we integrate out Eq. (\ref{bianchi-I-Einstein-4}) to have 
\begin{equation}\label{bianchi-I-Einstein-15}
\dot\sigma_1 = k \exp\left[-3\alpha_1 -\beta_1\right],
\end{equation}
where $k$ is a constant of integration. In addition, an initial condition for this equation can be figured out from the Friedmann equation (\ref{bianchi-I-Einstein-2}):
\begin{equation}\label{bianchi-I-Einstein-16}
\dot\alpha_0^2+\dot\alpha_0 \dot\beta_0 -H_1^2 =k^2 \exp\left[-6\alpha_0 -2\beta_0\right].
\end{equation}
As a result,  Eq. (\ref{bianchi-I-Einstein-15}) has a solution of $\sigma_1$:
\begin{eqnarray}\label{bianchi-I-Einstein-17}
\sigma_1 &=&\sigma_0 +\sqrt{\dot\alpha_0^2+\dot\alpha_0 \dot\beta_0 -H_1^2} \nonumber\\
&& \times \int \Biggl\{ \biggl[\cosh \left(3\tilde H_1 t\right)+\frac{\dot\alpha_0}{\tilde H_1} \sinh \left(3\tilde H_1 t \right) \biggr] \nonumber\\
&&  \times \biggl[\cosh \left(3\bar H_1 t\right)+\frac{\dot\beta_0}{3\bar H_1} \sinh \left(3\bar H_1 t \right) \biggr]\Biggr\}^{-1} dt, \nonumber\\
\end{eqnarray}
where $\sigma_0 =\sigma_1(t=0)$ as an initial value. 

In conclusion, we have found a set anisotropic solutions (\ref{bianchi-I-Einstein-12}), (\ref{bianchi-I-Einstein-14}), and (\ref{bianchi-I-Einstein-17}) for the five-dimensional Einstein field equations with the effective cosmological constant $\Lambda_M$ coming from the massive graviton terms ${\cal L}_i$ ($i=2-5$). 
\subsection{Stability analysis of Bianchi type I expanding solutions}
In this subsection, we would like to study the stability of this set of anisotropic solutions by considering the exponential perturbations ~\cite{WFK}:
\begin{eqnarray}
\delta \alpha_1 &=& {\cal C}_{\alpha} \exp[\kappa t]; ~\delta \sigma_1 = {\cal C}_{\sigma} \exp[\kappa t]; 
\delta \beta_1 = {\cal C}_{\beta} \exp[\kappa t]; \nonumber\\
\delta A &=& {\cal C}_{A} \exp[\kappa t]; ~\delta B = {\cal C}_{B} \exp[\kappa t]; ~\delta C = {\cal C}_{C} \exp[\kappa t];\nonumber\\
\delta N_2 &=& {\cal C}_{N_2} \exp[\kappa t].
\end{eqnarray}
As a result, a set of perturbation equations obtained by perturbing Eqs. (\ref{bianchi-I-Einstein-2}), (\ref{bianchi-I-Einstein-15}), (\ref{bianchi-I-Einstein-5}), (\ref{bianchi-constraint-1}), (\ref{bianchi-constraint-2}), (\ref{bianchi-constraint-3}), and (\ref{bianchi-constraint-4}) are
\begin{widetext}
\begin{equation}\label{matrix}
{\cal D} \left( {\begin{array}{*{20}c}
    {\cal C}_{\alpha} \\
    {\cal C}_{\sigma}   \\
    {\cal C}_{\beta} \\
    {\cal C}_{A} \\
    {\cal C}_{B}  \\
    {\cal C}_{C} \\
    {\cal C}_{N_2} \\
 \end{array} } \right) \equiv \left[ {\begin{array}{*{20}c}
   {A_{11} } & {A_{12} } & {A_{13} } & {0}  &{0} & {0} &{0} \\
   {A_{21} } & {A_{22} } & {A_{23} } & {0}& {0} &{0}&{0}\\
   {A_{31} } & {A_{32} } & {A_{33} } & {0}& {0} &{0}&{0}\\
   {A_{41} } & {A_{42} } & {A_{43} }  & {A_{44}} & {A_{45}}&{A_{46}}&{A_{47}}\\
   {A_{51} } & {A_{52} } & {A_{53}}  & {A_{54}} & {A_{55}} &{A_{56}}&{A_{57}}\\
   {A_{61} } & {A_{62} } & {A_{63} }  & {A_{64}} & {A_{65}}&{A_{66}}&{A_{67}} \\
   {A_{71} } & {A_{72} } & {A_{73} }  & {A_{74}} & {A_{75}}&{A_{76}}&{A_{77}} \\
 \end{array} } \right]\left( {\begin{array}{*{20}c}
    {\cal C}_{\alpha} \\
    {\cal C}_{\sigma}   \\
    {\cal C}_{\beta} \\
    {\cal C}_{A} \\
    {\cal C}_{B}  \\
    {\cal C}_{C} \\
    {\cal C}_{N_2} \\
 \end{array} } \right) = 0.
\end{equation}
\end{widetext}
It appears that the components $A_{ij}$ with $i=4-7$ and $j=1-7$ do not involve any variable $\kappa$ since Eqs.  (\ref{bianchi-constraint-1}), (\ref{bianchi-constraint-2}), (\ref{bianchi-constraint-3}), and (\ref{bianchi-constraint-4}) do not include any time derivative of $A$, $B$, $C$, and $N_2$. Hence, the components $A_{ij}$ with $i=4-7$ and $j=1-7$ do not affect on the stability property of the expanding solutions.  Therefore, we will consider the reduced matrix equation instead of Eq. (\ref{matrix}) in order to investigate whether the  set anisotropic solutions (\ref{bianchi-I-Einstein-12}), (\ref{bianchi-I-Einstein-14}), and (\ref{bianchi-I-Einstein-17}) is stable or not against the field perturbations:
\begin{equation}\label{matrix-reduced}
{\tilde {\cal D}} \left( {\begin{array}{*{20}c}
    {\cal C}_{\alpha} \\
    {\cal C}_{\sigma}   \\
    {\cal C}_{\beta} \\
 \end{array} } \right) \equiv \left[ {\begin{array}{*{20}c}
   {A_{11} } & {A_{12} } & {A_{13} } \\
   {A_{21} } & {A_{22} } & {A_{23} } \\
   {A_{31} } & {A_{32} } & {A_{33} } \\
 \end{array} } \right]\left( {\begin{array}{*{20}c}
    {\cal C}_{\alpha} \\
    {\cal C}_{\sigma}   \\
    {\cal C}_{\beta} \\
 \end{array} } \right) = 0,
\end{equation}
where 
\begin{equation}
A_{11} = \left(2\dot\alpha_1 +\dot\beta_1\right)\kappa; ~A_{12}=-2 \dot\sigma_1 \kappa; 
~A_{13}=\dot\alpha_1 \kappa;
\end{equation}
\begin{eqnarray}
A_{21} &=& 3\dot\sigma_1; ~A_{22}= \kappa; ~A_{23}=\dot\sigma_1 ; \\
A_{31}&=&-\left(4\dot\alpha_1 -\dot\beta_1 \right)\kappa; ~A_{32}=4\dot\sigma_1 \kappa; \\
 A_{33} &=& \kappa^2 +\left(\dot\alpha_1 +2\dot\beta_1\right)\kappa.
\end{eqnarray}
It is known that Eq. (\ref{matrix-reduced}) admits non-trivial solutions only when
\begin{equation}
\det {\tilde {\cal D}}=0,
\end{equation}
which can be calculated to be an equation of $\kappa$:
\begin{eqnarray}
\kappa\left[\left(3\dot\alpha_1^2 +2\dot\alpha_1 \dot\beta_1 +\dot\beta_1^2 +3\dot\sigma_1^2 \right)\kappa +3 \dot\sigma_1^2 \left(3\dot\alpha_1 +\dot\beta_1 \right)\right]=0. \nonumber\\
\end{eqnarray}
As a result, besides the trivial solution, $\kappa_1=0$, this equation implies the non-trivial solution:
\begin{equation}
\kappa_2 = -\frac{3 \dot\sigma_1^2 \left(3\dot\alpha_1 +\dot\beta_1 \right)}{3\dot\alpha_1^2 +2\dot\alpha_1 \dot\beta_1 +\dot\beta_1^2 +3\dot\sigma_1^2 }.
\end{equation}
It turns out that $\kappa_2<0$ for expanding cosmological solutions with $3\dot\alpha_1+\dot\beta_1 >0$. Hence, we conclude that the Bianchi type I expanding  solutions of five-dimensional nonlinear massive gravity are indeed stable against the field perturbations. Note again that the Bianchi type I expanding  solutions of four-dimensional nonlinear massive gravity have also been found and shown to be stable ~\cite{bianchi-I,WFK}. 
\section{Schwarzschild-Tangherlini black holes} \label{sec6}
In order to seek the five-dimensional Schwarzschild-Tangherlini black hole ~\cite{5d-sch,5d-sch-stability,5d-review} for the dRGT nonlinear massive gravity theory, we consider the following spherically symmetric metrics:
\begin{eqnarray} \label{five-dim-metric}
 g^{5d}_{\mu\nu}dx^{\mu}dx^{\nu} &= & -N_1^2\left(t,r\right) dt^2 +\frac{dr^2}{F_1^2\left(t,r\right)} \nonumber\\
&& +2D_1\left(t,r\right)dt dr+\frac{r^2 d\Omega_3^2}{H_1^2\left(t,r\right)},
\end{eqnarray}
\begin{eqnarray}
 Z^{5d}_{\mu\nu}dx^{\mu}dx^{\nu}&= & -N_2^2\left(t,r\right) dt^2 +\frac{dr^2}{F_2^2\left(t,r\right)} \nonumber\\
&& +2D_2\left(t,r\right)dt dr+\frac{r^2 d\Omega_3^2}{H_2^2\left(t,r\right)}, 
\end{eqnarray}
with
\begin{equation}
d\Omega_3^2=d\theta^2 +\sin^2 \theta d\varphi^2 + \sin^2 \theta \sin^2 \varphi d \psi^2.
\end{equation}
Here $(r,\theta,\varphi,\psi)$ are the spherical coordinates with allowed ranges: $0\leq \theta \leq \pi$, $0\leq \varphi \leq \pi$, and $0\leq \psi \leq 2\pi$.  In addition, $N_i(t,r)$, $F_i(t,r)$, $D_i(t,r)$, and $H_i(t,r)$ ($i=1-2$) are arbitrary functions of time $t$ and radial coordinate $r$~\cite{KNT,RGT2,RGT1}. Moreover, the  St\"uckelberg scalar fields have been chosen to be in the unitary gauge, $\phi^a =x^a$ ~\cite{KNT,RGT2,TMN,RGT1}.

As a result, the non-vanishing components of ${\cal K}^{\mu}{ }_\nu$ are defined to be
\begin{eqnarray} \label{component-K}
{\cal K}^0{ }_0 &=&1- \sqrt{\frac{N_2^2+D_1 D_2 F_1^2}{N_1^2 + D_1^2 F_1^2}}, \nonumber\\
 {\cal K}^0{ }_1 &=& -\frac{1}{F_2}\sqrt{\frac{D_1 F_1^2 -D_2 F_2^2}{ N_1^2 + D_1^2 F_1^2 }}, \nonumber\\ 
  {\cal K}^1{ }_0 &= &- F_1 \sqrt{\frac{N_1^2 D_2 -N_2^2 D_1}{N_1^2 + D_1^2 F_1^2}}, \nonumber\\
 {\cal K}^1{ }_1 &=& 1- \frac{F_1}{F_2}\sqrt{\frac{N_1^2 +D_1 D_2 F_2^2}{N_1^2 +D_1^2 F_1^2}}, \nonumber\\
 {\cal K}^2{ }_2&= &{\cal K}^3{ }_3 = {\cal K}^4{ }_4=1-\frac{H_1}{H_2}.
\end{eqnarray}
For convenience, we define some additional results: 
\begin{align} 
 [{\cal K}]^n =& \left( {\cal K}^0{ }_0 + {\cal K}^1{ }_1 +3 {\cal K}^2{ }_2 \right)^n, \nonumber\\
 [{\cal K}^2] =& \left({\cal K}^0{ }_0 \right)^2 +\left({\cal K}^1{ }_1 \right)^2+3 \left({\cal K}^2{ }_2\right)^2+2{\cal K}^0{ }_1  {\cal K}^1{ }_0, \nonumber\\
 [{\cal K}^3] =& \left({\cal K}^0{ }_0\right)^3 +\left({\cal K}^1{ }_1 \right)^3+3\left({\cal K}^2{ }_2\right)^3 \nonumber\\
&+3{\cal K}^0{ }_1  {\cal K}^1{ }_0 \left({\cal K}^0{ }_0+{\cal K}^1{ }_1\right), \nonumber\\
 [{\cal K}^4] = & \left({\cal K}^0{ }_0\right)^4 + \left({\cal K}^1{ }_1 \right)^4 + 3\left({\cal K}^2{ }_2\right)^4 + 4{\cal K}^0{ }_1  {\cal K}^1{ }_0 \nonumber\\
&\times\left[ \left({\cal K}^0{ }_0\right)^2 +{\cal K}^0{ }_0 {\cal K}^1{ }_1+ \left({\cal K}^1{ }_1\right)^2 +\frac{1}{2}{\cal K}^0{ }_1  {\cal K}^1{ }_0 \right] , \nonumber\\
  [{\cal K}^5] =& \left({\cal K}^0{ }_0\right)^5 + \left({\cal K}^1{ }_1 \right)^5 + 3\left({\cal K}^2{ }_2\right)^5 +5{\cal K}^0{ }_1  {\cal K}^1{ }_0 \nonumber\\
& \times \biggl[ \left({\cal K}^0{ }_0\right)^3 +\left({\cal K}^1{ }_1\right)^3+ \left({\cal K}^0{ }_0{\cal K}^1{ }_1+{\cal K}^0{ }_1  {\cal K}^1{ }_0\right)  \nonumber\\
 &  \times \left({\cal K}^0{ }_0 +{\cal K}^1{ }_1\right) \biggr]   .
\end{align}
Given the above results, the massive graviton terms ${\cal L}_{i}$'s ($i=2-5$)  are explicitly defined to be
\begin{eqnarray}
{\cal L}_2 &=& 2 \Bigl[ {\cal K}^0{ }_0 \left({\cal K}^1{ }_1 +3 {\cal K}^2{ }_2\right)+3 {\cal K}^2{ }_2 \left({\cal K}^1{ }_1 +{\cal K}^2{ }_2\right)  \nonumber\\
&& -{\cal K}^0{ }_1 {\cal K}^1{ }_0 \Bigr], \\
{\cal L}_3 &=& 2 {\cal K}^2{ }_2 \Bigl[  3{\cal K}^0{ }_0 \left({\cal K}^1{ }_1 + {\cal K}^2{ }_2\right)+{\cal K}^2{ }_2 \left(3{\cal K}^1{ }_1 +{\cal K}^2{ }_2 \right)  \nonumber\\
&&  -3{\cal K}^0{ }_1 {\cal K}^1{ }_0  \Bigr],\\
{\cal L}_4 &=& 2\left({\cal K}^2{ }_2\right)^2 \Bigl[{\cal K}^0{ }_0 \left( 3{\cal K}^1{ }_1 + {\cal K}^2{ }_2\right) +{\cal K}^1{ }_1{\cal K}^2{ }_2  \nonumber\\
&&  -3 {\cal K}^0{ }_1 {\cal K}^1{ }_0 \Bigr], \\
{\cal L}_5 &=&2 \left({\cal K}^2{ }_2\right)^3 \left({\cal K}^0{ }_0 {\cal K}^1{ }_1 -{\cal K}^0{ }_1 {\cal K}^1{ }_0  \right) .
\end{eqnarray}
Hence, the corresponding graviton Lagrangian ${\cal L}_M$ turns out to be
\begin{eqnarray} \label{Lagra-reduced}
{\cal L}_M &=&2\left\{ \left[\alpha_5 \left({\cal K}^2{ }_2\right)^3+ 3\alpha_4 \left({\cal K}^2{ }_2\right)^2 +3 \alpha_3 {\cal K}^2{ }_2 +1\right] \right. \nonumber\\
&& \left. \times \left({\cal K}^0{ }_0 {\cal K}^1{ }_1 - {\cal K}^0{ }_1 {\cal K}^1{ }_0\right) \right.\nonumber\\ 
&& \left.+{\cal K}^2{ }_2 \left[ \alpha_4 \left({\cal K}^2{ }_2\right)^2+3\alpha_3 {\cal K}^2{ }_2 +3\right] \left({\cal K}^0{ }_0 + {\cal K}^1{ }_1\right) \right.\nonumber\\
&&\left.+ \left({\cal K}^2{ }_2\right)^2 \left(\alpha_3{\cal K}^2{ }_2+3 \right)\right\}.
\end{eqnarray}

According to Refs.~\cite{KNT,RGT2}, there exists a constraint equation, which is associated with the non-diagonal component of physical metric, $g_{0r}$, given by
\begin{equation}
g_{0r}R_{00}-g_{00}R_{0r}=0.
\end{equation} 
Furthermore, thanks to the Einstein field equations (\ref{Einstein-5d}) in vacuum ($T_{\mu\nu}=0$)  this equation can be reduced to 
\begin{equation}
g_{0r} X_{00} -g_{00}X_{0r}+\sigma \left( g_{0r}Y_{00}- g_{00} Y_{0r}\right) =0.
\end{equation}
As a result, this equation can be defined to give an equation:
\begin{equation} \label{constraint-equation-BH}
  {\cal K}^0{ }_1 N_1^4 + \left({\cal K}^0{ }_0 -{\cal K}^1{ }_1\right) D_1 N_1^2 -{\cal K}^1{ }_0 D_1^2 =0,
\end{equation}
which will be fulfilled if $D_1 =D_2=0$, consistent with investigations in Refs.~\cite{KNT,RGT2}. Note that if either $D_1$ or $D_2$ is set to be zero then the corresponding $D_2$ or $D_1$ must also be zero due to the requirement that both ${\cal K}^0{ }_1$ and ${\cal K}^1{ }_0$ shown in Eq. (\ref{component-K}) must be real definite.  Note also that  Eq. (\ref{constraint-equation-BH}) might admit non-vanishing $D_1$ and $D_2$ solutions, which correspond to the non-diagonal black hole metrics ~\cite{non-bidiagonal-BH}. For example, choosing solutions such as
\begin{equation}
\frac{N_1^2}{N_2^2} = \frac{D_1}{D_2} =\frac{F_2^2}{F_1^2}
\end{equation}
will lead to 
\begin{eqnarray}
{\cal K}^0{ }_1 &=& {\cal K}^1{ }_0 =0 ,\\
{\cal K}^0{ }_0 &=& {\cal K}^1{ }_1,
\end{eqnarray}
resulting the vanishing of the left-hand side of Eq. (\ref{constraint-equation-BH}). Furthermore, if we choose a case that
\begin{equation}
\frac{N_1^2}{N_2^2} = \frac{D_1}{D_2} =\frac{F_2^2}{F_1^2} =\frac{H_2^2}{H_1^2} = {\cal C},
\end{equation}
which corresponds to a case of proportional metrics, i.e., $f_{\mu\nu} = {\cal C} g_{\mu\nu}$, then we would obtain the following non-diagonal black hole metrics ~\cite{non-bidiagonal-BH} for the five-dimensional nonlinear massive gravity.

However, we will only consider the case of $D_1 =D_2 =0$, which leads to ${\cal K}^0{ }_1 ={\cal K}^1{ }_0=0$ and
\begin{eqnarray}
{\cal K}^0{ }_0 &=& 1-\frac{N_2}{N_1}, \\
{\cal K}^1{ }_1 &=&1 -\frac{F_1}{F_2},
\end{eqnarray}
for simplicity from now on.   In the next subsection, we will try to solve constraint equations associated with the non-vanishing components of the fiducial metric. In particular, once the constraint equations are solved analytically (or numerically), we can define the corresponding graviton terms $ {\cal L}_M \equiv {\cal L}_2 + \alpha_3 {\cal L}_3 + \alpha_4 {\cal L}_4 +\alpha_5 {\cal L}_5$ ~\cite{WFK}.  Indeed, we will be able to compute an effective cosmological constant associated with the graviton terms such as  $\Lambda_M= -m_g^2 {\cal L}_M/2$, provided that the fiducial metric is compatible with the physical metric, similar to the study  in Ref. ~\cite{WFK}. 
\subsection{Constraint equations}
Similar to the previous sections, we will derive and then solve constraint equations associated with the existence of non-trivial factors of the fiducial metric: $N_2$, $ F_2$, and $H_2$ in order to show constant-like behavior of graviton terms. Since  the massive graviton Lagrangian ${\cal L}_M$ shown in Eq. (\ref{Lagra-reduced}) does not contain any time or radial coordinate derivative of $N_2$, $ F_2$, and $H_2$, then we have the following variational equations:
\begin{eqnarray} \label{eq17}
\frac{\partial {\cal L}_M}{\partial N_2} =0;~\frac{\partial {\cal L}_M}{\partial F_2} =0; ~\frac{\partial {\cal L}_M}{\partial H_2} =0.
\end{eqnarray}
As a result, these variational equations can be rewritten to be
\begin{eqnarray}\label{eq34}
 \frac{\partial {\cal L}_M}{\partial N_2} &=&\frac{\partial {\cal L}_M}{\partial {\cal K}^0{ }_0} \frac{\partial{\cal K}^0{ }_0}{\partial N_2}  =0, \\
\label{eq36}
 \frac{\partial {\cal L}_M}{\partial F_2} &=&\frac{\partial {\cal L}_M}{\partial {\cal K}^1{ }_1} \frac{\partial {\cal K}^1{ }_1}{\partial F_2} =0,\\
\label{eq37}
 \frac{\partial {\cal L}_M}{\partial H_2}& =&\frac{\partial {\cal L}_M}{\partial {\cal K}^2{ }_2} \frac{\partial {\cal K}^2{ }_2}{\partial H_2}=0,
\end{eqnarray}
respectively. As a result, solutions to a set of Eqs. (\ref{eq34}), (\ref{eq36}), and (\ref{eq37}) are solved to be
\begin{eqnarray}\label{eq38}
\frac{\partial {\cal L}_M}{\partial {\cal K}^0{ }_0} = 0;~ \frac{\partial {\cal L}_M}{\partial {\cal K}^1{ }_1} =0; ~ \frac{\partial {\cal L}_M}{\partial {\cal K}^2{ }_2} =0.
\end{eqnarray}
As will be shown below, these equations will lead to simplified constraint equations, which seem to be more convenience to solve. 
By using the explicit definition of ${\cal L}_M $ shown in Eq. (\ref{Lagra-reduced}), we are able to define the following constraint equations:
\begin{widetext}
\begin{eqnarray} \label{constraint1}
\left(\alpha_5 {\cal K}^1{ }_1 +\alpha_4 \right)\left({\cal K}^2{ }_2\right)^3 +3 \left(\alpha_4 {\cal K}^1{ }_1 + \alpha_3 \right)\left({\cal K}^2{ }_2\right)^2 +3 \left(\alpha_3 {\cal K}^1{ }_1 +1 \right){\cal K}^2{ }_2+{\cal K}^1{ }_1 &=&0, \\
\label{constraint4}
\left(\alpha_5 {\cal K}^0{ }_0 +\alpha_4 \right)\left({\cal K}^2{ }_2\right)^3 +3 \left(\alpha_4 {\cal K}^0{ }_0 + \alpha_3 \right)\left({\cal K}^2{ }_2\right)^2 +3 \left(\alpha_3 {\cal K}^0{ }_0 +1 \right){\cal K}^2{ }_2+{\cal K}^0{ }_0 &=&0, \\
\label{constraint5}
 \left[\alpha_5 \left({\cal K}^2{ }_2\right)^2 +2\alpha_4 {\cal K}^2{ }_2 +\alpha_3 \right] {\cal K}^0{ }_0 {\cal K}^1{ }_1+ \left[\alpha_4 \left({\cal K}^2{ }_2\right)^2 +2 \alpha_3 {\cal K}^2{ }_2 +1 \right] \left({\cal K}^0{ }_0+{\cal K}^1{ }_1\right)+ \left(\alpha_3 {\cal K}^2{ }_2+2\right){\cal K}^2{ }_2 &=&0.
\end{eqnarray}
\end{widetext}
Now, we are going to solve these constraint equations. 
First, combining both Eqs. (\ref{constraint1}) and (\ref{constraint4}) leads to
\begin{eqnarray} \label{constraint1-4}
 \left[ \alpha_5 \left({\cal K}^2{ }_2\right)^3+ 3\alpha_4 \left({\cal K}^2{ }_2\right)^2 +3 \alpha_3 {\cal K}^2{ }_2 +1\right] && \nonumber\\
 \times \left({\cal K}^0{ }_0 -{\cal K}^1{ }_1\right)=0&&,
\end{eqnarray}
which implies two possible cases:
\begin{eqnarray}
&&{\cal K}^0{ }_0 = {\cal K}^1{ }_1, \\
\label{case2}
&& \alpha_5 \left({\cal K}^2{ }_2\right)^3+ 3\alpha_4 \left({\cal K}^2{ }_2\right)^2 +3 \alpha_3 {\cal K}^2{ }_2 +1 =0.
\end{eqnarray}
Below, we will discuss whether these solutions lead to physical solutions of the constraint equations.
\subsubsection{Case 1}
In this case, we will consider the first solution of Eq. (\ref{constraint1-4}), ${\cal K}^0{ }_0 = {\cal K}^1{ }_1$. Thanks to this solution,  Eq. (\ref{constraint5}) can be reduced to
\begin{eqnarray} \label{constraint6}
&&\left[\alpha_5 \left({\cal K}^2{ }_2\right)^2 +2\alpha_4 {\cal K}^2{ }_2 +\alpha_3 \right] \left({\cal K}^1{ }_1\right)^2 \nonumber\\
&&+2 \left[\alpha_4 \left({\cal K}^2{ }_2\right)^2 +2 \alpha_3 {\cal K}^2{ }_2 +1 \right] {\cal K}^1{ }_1 \nonumber\\
&& + \left(\alpha_3 {\cal K}^2{ }_2+2\right) {\cal K}^2{ }_2=0. 
\end{eqnarray}
Furthermore, combining both Eqs. (\ref{constraint1}) and (\ref{constraint6}) gives an equation:
\begin{eqnarray}
&&\left({\cal K}^1{ }_1 - {\cal K}^2{ }_2\right) \left\{  \left[\alpha_5 \left({\cal K}^2{ }_2\right)^2 +2\alpha_4 {\cal K}^2{ }_2 +\alpha_3 \right]{\cal K}^1{ }_1 \right. \nonumber\\
&& \left.+ \alpha_4 \left({\cal K}^2{ }_2\right)^2 +2\alpha_3{\cal K}^2{ }_2 +1  \right\}=0,
\end{eqnarray}
which is solved to give
\begin{eqnarray}\label{solution-a}
{\cal K}^1{ }_1 &= &{\cal K}^2{ }_2 , \\
\label{solution-b}
{\cal K}^1{ }_1 &=& -\frac{\alpha_4 \left({\cal K}^2{ }_2\right)^2 +2\alpha_3{\cal K}^2{ }_2 +1}{\alpha_5 \left({\cal K}^2{ }_2\right)^2 +2\alpha_4 {\cal K}^2{ }_2 +\alpha_3}.
\end{eqnarray}

Now, for the first solution (\ref{solution-a}) the equation (\ref{constraint1}) or (\ref{constraint5}) is reduced to
\begin{equation} \label{solution-c}
\alpha_5 \left({\cal K}^2{ }_2\right)^3 + 4 \alpha_4 \left({\cal K}^2{ }_2\right)^2 +6\alpha_3 {\cal K}^2{ }_2 +4 =0,
\end{equation}
here we have ignored the trivial solution, ${\cal K}^2{ }_2=0$, which leads to the vanishing graviton terms. 

Given the solutions, ${\cal K}^0{ }_1= {\cal K}^1{ }_0=0$ and ${\cal K}^0{ }_0={\cal K}^1{ }_1={\cal K}^2{ }_2$, we can define the corresponding value of graviton terms ${\cal L}_M$ to be
\begin{equation}\label{case1-LM}
{\cal L}_M = 2 \left({\cal K}^2{ }_2\right)^2 \left[\alpha_4 \left({\cal K}^2{ }_2\right)^2 +4 \alpha_3 {\cal K}^2{ }_2 +6\right],
\end{equation}
here the value of ${\cal K}^2{ }_2$   can be determined in terms of the field parameters $\alpha_{3}$, $\alpha_4$, and $\alpha_5$ thanks to the constraint equation  (\ref{solution-c}). Note that the massive graviton terms will play as a real effective cosmological constant, 
\begin{equation} 
\Lambda_M=  -m_g^2 \left({\cal K}^2{ }_2\right)^2 \left[\alpha_4 \left({\cal K}^2{ }_2\right)^2 +4 \alpha_3 {\cal K}^2{ }_2 +6\right] .
\end{equation}
 
Next, we will discuss the second solution shown in Eq. (\ref{solution-b}). As a result, from Eq. (\ref{constraint1}) we obtain
\begin{equation} \label{case1-eqa}
{\cal K}^1{ }_1 = -\frac{{\cal K}^2{ }_2 \left(\alpha_3 {\cal K}^2{ }_2 +2\right)}{\alpha_4 \left({\cal K}^2{ }_2\right)^2 +2\alpha_3 {\cal K}^2{ }_2+1}.
\end{equation}
By combining this result with the equation (\ref{solution-b}), we arrive at a quartic function of $ {\cal K}^2{ }_2$ as
\begin{eqnarray} \label{case1-eqb}
&&\left(\alpha_3 \alpha_5 -\alpha_4^2\right) \left( {\cal K}^2{ }_2\right)^4 +2 \left(\alpha_5 -\alpha_3 \alpha_4 \right)\left( {\cal K}^2{ }_2\right)^3 \nonumber\\
&& + \left(2\alpha_4 -3\alpha_3^2 \right)\left( {\cal K}^2{ }_2\right)^2 -2\alpha_3  {\cal K}^2{ }_2 -1 =0.
\end{eqnarray}
It is known that this quartic equation admits four  roots  $ {\cal K}^2{ }_2$, which can be either real or complex definite. Once the value of $ {\cal K}^2{ }_2$ is determined, the value of ${\cal K}^1{ }_1$ can be evaluated consistently, following Eq. (\ref{case1-eqa}).
\subsubsection{Case 2}
In this case, the second solution of Eq. (\ref{constraint1-4}),  $\alpha_5 \left({\cal K}^2{ }_2\right)^3+ 3\alpha_4 \left({\cal K}^2{ }_2\right)^2 +3 \alpha_3 {\cal K}^2{ }_2 +1 =0$, will be used. As a result, thanks to this equation both constraint equations (\ref{constraint1}) and (\ref{constraint4}) are reduced to
\begin{equation} \label{case2-equ-1}
\alpha_4 \left({\cal K}^2{ }_2\right)^2+3\alpha_3 {\cal K}^2{ }_2 +3 =0,
\end{equation}
here  the trivial solution, ${\cal K}^2{ }_2=0$ resulting zero graviton terms, has been neglected.  As a result, the non-trivial solutions of Eq. (\ref{case2-equ-1}) are solved to be
\begin{equation}
{\cal K}^2{ }_2 = \frac{-3\alpha_3 \pm \sqrt{9\alpha_3^2 -12\alpha_4}}{2\alpha_4}
\end{equation}
along with the relation between the parameter $\alpha_5$ and the other $\alpha_{3,4}$  such as
\begin{eqnarray}
\alpha_5 &=&- \frac{3\alpha_4 \left({\cal K}^2{ }_2\right)^2 +3 \alpha_3 {\cal K}^2{ }_2 +1}{\left({\cal K}^2{ }_2\right)^3} \nonumber\\
&=& \frac{8\alpha _4^2 \left[\left(9 \alpha _3^2-8 \alpha _4\right)\mp 3 \alpha _3 \sqrt{9 \alpha _3^2-12 \alpha _4}  \right] }{\left(3 \alpha _3 \mp \sqrt{9 \alpha _3^2-12 \alpha _4}\right)^3}, \nonumber\\
\end{eqnarray}
which is identical to that defined in Eq. (\ref{special-a5}).
Consequently, the corresponding graviton terms turn out to be
\begin{eqnarray}
{\cal L}_M &=& 2 \left({\cal K}^2{ }_2\right)^2 \left({\alpha_3 \cal K}^2{ }_2+3\right)\nonumber\\
&=& -\frac{3}{\alpha_4^3} \biggl[9\alpha_3^4 -18\alpha_3^2 \alpha_4 +6\alpha_4^2  \nonumber\\
&&  \mp \alpha_3 \left(3\alpha_3 ^2-4\alpha_4 \right) \sqrt{9\alpha_3^2 -12\alpha_4}\biggr],
\end{eqnarray}
which is independent of values of  ${\cal K}^0{ }_0$ and $ {\cal K}^1{ }_1$. Hence, the corresponding effective cosmological constant can be defined to be
\begin{eqnarray}
\Lambda_M &=&  \frac{3m_g^2}{2\alpha_4^3} \biggl[9\alpha_3^4 -18\alpha_3^2 \alpha_4 +6\alpha_4^2  \nonumber\\
 &&  \mp \alpha_3 \left(3\alpha_3 ^2-4\alpha_4 \right) \sqrt{9\alpha_3^2 -12\alpha_4}\biggr],
\end{eqnarray}
here the field parameters $\alpha_3$ and $\alpha_4$ are assumed to satisfy the condition such that $\alpha_3^2>4\alpha_4/3$. Moreover, the effective cosmological constant $\Lambda_M$, which is independent of $\alpha_5$,  can be either negative or positive definite, depending on the value of $\alpha_3$ and $\alpha_4$.

Additionally, these variables ${\cal K}^0{ }_0$ and $ {\cal K}^1{ }_1$ seem to obey the following constraint equation, which comes from Eq. (\ref{constraint5}),
\begin{eqnarray} \label{constraint7}
&& \left(\alpha_3 {\cal K}^2{ }_2+2\right) {\cal W} =0, \\
&& {\cal W} \equiv \left[ {\cal K}^2{ }_2 \left( {\cal K}^2{ }_2-{\cal K}^0{ }_0-{\cal K}^1{ }_1 \right) +{\cal K}^0{ }_0 {\cal K}^1{ }_1 \right],
\end{eqnarray}
here we have used the results shown in Eqs. (\ref{case2}) and (\ref{case2-equ-1}) in order to simplify the  Eq. (\ref{constraint5}). It appears that we only have one equation for three independent variables. Hence, we cannot define explicit values of ${\cal K}^0{ }_0$ and $ {\cal K}^1{ }_1$  in general.  It is apparent that if ${\cal K}^2{ }_2 \neq -2/{\alpha_3}$ along with an assumption, ${\cal K}^0{ }_0={\cal K}^1{ }_1$,  then
\begin{equation}\label{case2-sol-1}
{\cal K}^0{ }_0={\cal K}^1{ }_1 = {\cal K}^2{ }_2= \frac{-3\alpha_3 \pm \sqrt{9\alpha_3^2 -12\alpha_4}}{2\alpha_4}.
\end{equation}

On the other hand, it turns out that Eq. (\ref{constraint7}) always admits a solution for arbitrary ${\cal K}^0{ }_0$ and $ {\cal K}^1{ }_1$ satisfying the condition ${\cal W}\neq 0$:
\begin{equation} \label{case2-sol-2}
{\cal K}^2{ }_2 =-\frac{2}{\alpha_3}, 
\end{equation}
which implies the following relations between $\alpha_{3,4,5}$ as
\begin{equation}\label{case2-sol-3}
\alpha_4 = \frac{3 \alpha_3^2}{4}; ~ \alpha_5 =\frac{\alpha_3^3}{2}
\end{equation}
along with the negative cosmological constant:
\begin{equation}\label{case2-LM}
\Lambda_M = -m_g^2 \frac{{\cal L}_M} {2} =- \frac{4m_g^2}{\alpha_3^2} <0.
\end{equation}
More interestingly, for the relation in Eq. (\ref{case2-sol-3}) the solution in Eq. (\ref{case2-sol-1}) becomes
\begin{equation}
{\cal K}^0{ }_0={\cal K}^1{ }_1 = {\cal K}^2{ }_2=- \frac{2}{\alpha_3}.
\end{equation}
It is apparent that the solutions shown in Eqs. (\ref{case2-sol-2}) and (\ref{case2-sol-3}) satisfy both Eqs. (\ref{solution-c}) and (\ref{case1-eqb}) in the case 1, where ${\cal K}^0{ }_0={\cal K}^1{ }_1$. Additionally, the corresponding graviton term ${\cal L}_M$ in Eq. (\ref{case1-LM}) also takes the same value as shown in Eq. (\ref{case2-LM}), i.e., ${\cal L}_M = 8/\alpha_3^2 >0$. Hence, we conclude that the solutions shown in Eqs. (\ref{case2-sol-2}) and (\ref{case2-sol-3}) along with the corresponding effective cosmological constant $\Lambda_M \equiv -m_g^2 {\cal L}_M/ {2} =- {4m_g^2}/{\alpha_3^2}$ hold for every components of both physical and fiducial metrics. Note that this result is also valid in the four-dimensional nonlinear massive gravity assuming the physical metric is compatible with the fiducial one. Therefore, it seems that this fact is a unique feature of the dRGT nonlinear massive gravity, which could be true in both higher and lower dimensional scenarios. 
\subsection{Einstein field equations and their black hole solutions}
It is apparent that the five-dimensional Einstein field equations (\ref{Einstein-5d-reduced}) can be rewritten, thanks to the constant-like behavior of graviton terms as shown in the previous subsection, as follows
\begin{equation} \label{blackhole-Einstein-1}
\left({R_{\mu\nu}-\frac{1}{2}Rg_{\mu\nu}}\right) +\Lambda_M g_{\mu\nu}=0,
\end{equation}
here $\Lambda_M \equiv -m_g^2 {\cal L}_M/2$, the effective cosmological constant, can be either negative or positive definite depending on the values of field paramters $\alpha_i$ ($i=3-5$).

It has been shown that the five-dimensional Einstein equations with a cosmological constant $\Lambda_M$ admit the Schwarzschild-Tangherlini-(anti-)de Sitter black holes as their solutions ~\cite{5d-sch,5d-sch-stability}. In particular, the five-dimensional physical metric (\ref{five-dim-metric}) can be solved to be ~\cite{5d-sch,5d-sch-stability}
\begin{equation} \label{five-dim-metric-1}
ds^2 = -f(r)dt^2 +\frac{dr^2}{f(r)}+r^2d\Omega_3^2,
\end{equation}
with
\begin{eqnarray}
N_1^2(t,r)&=&F_1^2(t,r)=f(r)=1-\frac{\mu}{r^2}-\frac{\Lambda_M}{6}r^2,\\
H_1^2(t,r)&=&1, ~D_1^2(t,r)=0.
\end{eqnarray}
In addition, $\mu$ is a mass parameter defined as
\begin{equation}
\mu =\frac{8 G_5 M}{3\pi},
\end{equation}
where $M$ stands for the mass of source and $G_5$ denotes the 5-dimensional Newton constant. Furthermore, if $\Lambda_M$ is  positive definite  or negative definite then the metric (\ref{five-dim-metric-1}) will be called the  Schwarzschild-Tangherlini-de Sitter or Schwarzschild-Tangherlini-anti-de Sitter black hole, respectively; otherwise $\Lambda_M=0$ corresponds to the pure Schwarzschild-Tangherlini black hole. It is worth noting that the stability of the  Schwarzschild-Tangherlini-A(dS) black holes for massless gravitons against linearized gravitational perturbations has been analyzed in details in Ref. ~\cite{5d-sch-stability}. As a result, all these black holes have been proved to be  stable against linearized gravitational perturbations ~\cite{5d-sch-stability}. In the massive gravity, however, the stability of its corresponding Schwarzschild-Tangherlini-A(dS) black holes shown above might be  modified due to the existence of mass of graviton, which is small but non-zero. In particular, ones have shown in Ref. \cite{stability-BH} that the  Schwarzschild black holes in four-dimensional massive theory for both dynamical or non-dynamical fiducial metrics  turn out to be unstable against radial perturbations. Hence, one could expect that the Schwarzschild-Tangherlini-A(dS) black holes of five (or higher)-dimensional massive gravities are also unstable. Note that for the case of non-vanishing $D_1$ and $D_2$, it has been shown in Ref. \cite{non-bidiagonal-BH} that the corresponding  four-dimensional non-diagonal black holes seem to be stable against radial perturbations. Similarly, one could expect that the non-diagonal black hole metrics of five (or higher)-dimensional massive gravities also turn out to be stable.   Of course,  it needs further investigations to obtain correct confirmation because the physical and fiducial black hole metrics  in Ref. \cite{stability-BH} have been taken to be proportional to each other, while the physical and fiducial black hole metrics in the present paper both are  of the same form but not necessarily proportional to each other. Note that we should be aware of the existence of new graviton terms such as ${\cal L}_5$, which might affect on the stability of black holes of higher dimensional massive gravity theories. 
\section{Conclusions}\label{con}
Inspired by the success of the four-dimensional nonlinear massive gravity proposed by de Rham, Gabadadze, and Tolley (dRGT)  recently ~\cite{RGT}, we would like to construct higher dimensional graviton terms and study their cosmological implications. Our construction is based on the well-known Cayley-Hamilton theorem, which states that  any square matrix must obey its characteristic equation. As a result, we have been able to re-build up the four-dimensional graviton terms of the ghost-free dRGT theory by applying the characteristic equation for the matrix ${\cal K}^\mu{ }_\nu $. Furthermore,  several higher dimensional graviton terms, e.g., ${\cal L}_5$, ${\cal L}_6$, and ${\cal L}_7$ in five-, six-, and seven-dimensional spacetimes, respectively, have also been constructed consistently by following the same technique  used for the four-dimensional graviton terms. Additionally,  we have also shown that any ghost-like pathology arising at $m-$order levels ($m\geq n$) will no longer exist in the $n$-dimensional nonlinear massive gravity following the analysis used in Ref. ~\cite{RGT}.

For heuristic reasons, we have studied the cosmological implications of the five-dimensional nonlinear massive gravity with the additional graviton term, ${\cal L}_5$. In particular, several well-known metrics such as FLRW, Bianchi type I, and Schwarzschild-Tangherlini metrics have been shown to exist in the five-dimensional dRGT theory, thanks to the constant-like behavior of graviton terms under the assumption that the fiducial metric is compatible with the physical one. More interestingly, the stability analysis has been done to show that the Bianchi type I expanding solutions are stable against field perturbations. For the stability of the Schwarzschild-Tangherlini-(A)dS black holes, we might employ the investigation in Ref.~\cite{5d-sch-stability} since the massive graviton terms turn out to be nothing but an effective cosmological constant $\Lambda_M$ defined by $\Lambda_M \equiv -m_g^2 {\cal L}_M/2$. However, it is noted that  the stability analysis in Ref.~\cite{5d-sch-stability}, which have confirmed that Schwarzschild-Tangherlini-(A)dS black holes  are stable against linearized gravitational perturbations, deal only with massless gravitons. Hence, we should be very careful when investigating the stability black holes of massive gravity with non-vanishing graviton terms. For example, Ref. \cite{stability-BH} has shown that the  Schwarzschild black holes in four-dimensional massive theory are generally unstable against radial perturbations. One hence could expect the same result for the Schwarzschild-Tangherlini-(A)dS black holes of five-dimensional massive gravity, where the massive graviton term ${\cal L}_5$ has been introduced.

Additionally, we have shown that the value of four-dimensional effective cosmological constant $\Lambda^0_M$ defined by Eq. (\ref{lambda1}) can be recovered in the five-dimensional dRGT theory under a requirement that the additional parameter, $\alpha_5$, cannot be free but constrained by the other parameters, $\alpha_3$ and $\alpha_4$, as shown in Eq. (\ref{special-a5}).

Given the explicit expressions of ${\cal L}_6$ and ${\cal L}_7$ shown in Eqs. (\ref{L6}) and (\ref{L7}), respectively, one can similarly compute the following effective cosmological constants in six- and seven-dimensional nonlinear massive gravity for some particular metrics. Furthermore, one can also be able to figure out the following metrics for the higher dimensional dRGT theory by applying the same method presented in this paper. Other higher dimensional black holes ~\cite{5d-review}, e.g., the  Myers-Perry black holes, which are higher dimensional generalizations of Kerr black hole ~\cite{5d-kerr}, might be shown to exist in the higher dimensional scenarios of the nonlinear massive gravity in further investigations.

On the other hand, one can propose a specific five (or higher)-dimensional bigravity ~\cite{SFH,higher-bigravity} by mimicking graviton terms of five  (or higher)-dimensional nonlinear massive gravity. Note that the fiducial metric in the bigravity  acts as a dynamical metric, not non-dynamical as in the massive gravity. Hence, solutions of the bigravity might be  different from that of the massive gravity. In the massive gravity, our main task is to solve the constraint equations of the scale factors of fiducial metric, which could be algebraic equations if the unitary gauge of St\"uckelberg scalar fields is chosen. After solving the constraint equations, we will be able to calculate the following value of massive graviton terms, which will be identified as an effective cosmological constants $\Lambda_M \equiv -m_g^2 {\cal L}_M/2$. Hence, the Einstein field equations will be reduced to simple ones, in which all complicated graviton terms are hidden in the effective constant, $\Lambda_M$.  In the bigravity, however, the constraint equations coming from the fiducial sector will  no longer be algebraic but differential equations. Therefore, the value of massive graviton in the bigravity could not be easy to investigate, resulting the complexity of Einstein field equations. Hence,  higher dimensional bigravity models and their cosmological implications need further investigations, which would be presented elsewhere ~\cite{TQD}. For heuristic reasons, one can also extend some extensions of the dRGT theory such as  the quasi-dilaton model ~\cite{quasi} and its extension ~\cite{quasi-gen}, the mass-varying massive gravity ~\cite{varying}, the extended massive gravity ~\cite{extended}, the f(R) nonlinear massive gravity ~\cite{fr}, and the massive gravity with non-minimal coupling of matter ~\cite{non-minimal} to the higher dimensional scenarios proposed in this paper. Of course,  a full detailed confirmation for the ghost-free property of the higher dimensional nonlinear massive gravities also needs to be investigated systematically as for the four-dimensional scenario ~\cite{proof}. 

We hope that the study in the present paper could shed more light on the physics of massive gravity and its extensions.
\begin{acknowledgments}
The author would like to thank Profs. A. J. Tolley, C. de Rham, S. D. Odintsov, and S. Nojiri very much for their correspondence. The author highly appreciates Drs. L. Heisenberg, R. Brito,  E. Babichev, and S. Y.  Zhou  for their useful comments.  The author is deeply grateful to Prof. W. F.  Kao of Institute of Physics in National Chiao Tung Universtiy  for his useful advice  on massive gravity. This research is supported in part by  VNU University of Science, Vietnam National University, Hanoi.
\end{acknowledgments}

\end{document}